\documentclass[a4paper]{raa}

\usepackage{graphicx,times}
\usepackage[a4paper,textwidth=150mm,textheight=250mm,centering]{geometry}
\graphicspath{{figures/}}
\usepackage{natbib}
\usepackage{amssymb,amsmath}
\bibpunct{(}{)}{;}{a}{}{,}

\usepackage[scheme=plain,fontset=fandol]{ctex}

\usepackage[hidelinks]{hyperref}

\usepackage{ulem}
\usepackage{xspace}

\usepackage{etoolbox}
\makeatletter
\let\arxivOriginalMakeheadbox\makeheadbox
\patchcmd{\arxivOriginalMakeheadbox}{\hb@xt@ 40cc}{\hb@xt@\dimexpr\textwidth-8pt\relax}{}{\errmessage{Could not fit RAA heading}}
\renewcommand{\makeheadbox}{\begingroup
\let\if@referee\iftrue
\arxivOriginalMakeheadbox
\endgroup}
\patchcmd{\@maketitle}{\if@referee\makeheadbox\fi}{\makeheadbox}{}{\errmessage{Could not restore RAA heading}}
\makeatother

\begin{document}

  \title{Breaking the Chain: The Dynamical History of the Barnard’s Star Planetary System}

   \setcounter{page}{1}

   \author{Tian Yi (易天)
      \inst{1,*}\footnotetext{$*$Corresponding Author; yit23@mails.tsinghua.edu.cn}
   \and Qiran Huang (黄启然)
      \inst{2,3}
   \and Shuo Huang (黄硕)
      \inst{1,4}
   \and Chris W. Ormel
      \inst{1}
   }

   \institute{Department of Astronomy, Tsinghua University, Haidian DS 100084 Beijing, China; {\it yit23@mails.tsinghua.edu.cn}\\
        \and
             St. Paul's School, Concord, NH 03301, USA\
        \and
             Department of Computer Science and Engineering,
    University of California, San Diego, La Jolla, CA 92093, USA\\
        \and Tsung-Dao Lee Institute, Shanghai Jiao Tong University No.1 Lisuo Road, Pudong New Area, Shanghai, China\\
\vs\no
   }

\abstract{
Barnard's Star is the nearest single-star system to the Sun and hosts four recently detected sub-Earth-mass planets. The system exhibits architectural imprints of a ``broken resonant chain''.  We explore scenarios in which the presence of a now-lost, additional planet provides a more natural explanation for the present-day architecture from a dynamical-formation perspective. In such a situation, a resonance-breaking event triggered a merger between two planets, forming the present-day planet c. We find two viable pathways that can reproduce the observed configuration: (i) stellar-driven, long-term tidal expansion and (ii) disk-driven, early expansion, in which one-sided Lindblad torques separate the two outermost planets. Both scenarios can lead to the c1--c2 merger required to form the present-day architecture. {The latter pathway} also agrees with the broader exoplanet census: among the 14 compact four-planet systems we identify with broken-chain architectures similar to Barnard's Star system, ten have the third planet ``breaking'' the resonant sequence.
}

   \authorrunning{Yi, Huang, Huang and Ormel}
   \titlerunning{Dynamical History of Barnard's Star Planets}

   \maketitle

\section{Introduction}
\label{sect:intro}

Since the discovery of the first exoplanet, transit and radial-velocity surveys have revealed a diverse population of planetary systems. The NASA Exoplanet Archive currently lists more than 6300 confirmed exoplanets distributed among thousands of systems.\footnote{NASA Exoplanet Archive, \url{https://exoplanetarchive.ipac.caltech.edu}} Their host stars span a wide range of masses, ages, and distances. A particularly nearby example is Barnard's Star, whose compact system of four sub-Earth-mass planets was identified and confirmed through recent extreme-precision radial-velocity observations \citep{HernandezEtal2024, BasantEtal2025}. The system has since been examined in further dynamical studies \citep{ByrneEtal2026}.

\begin{figure}
\centering
\includegraphics[width=\textwidth]{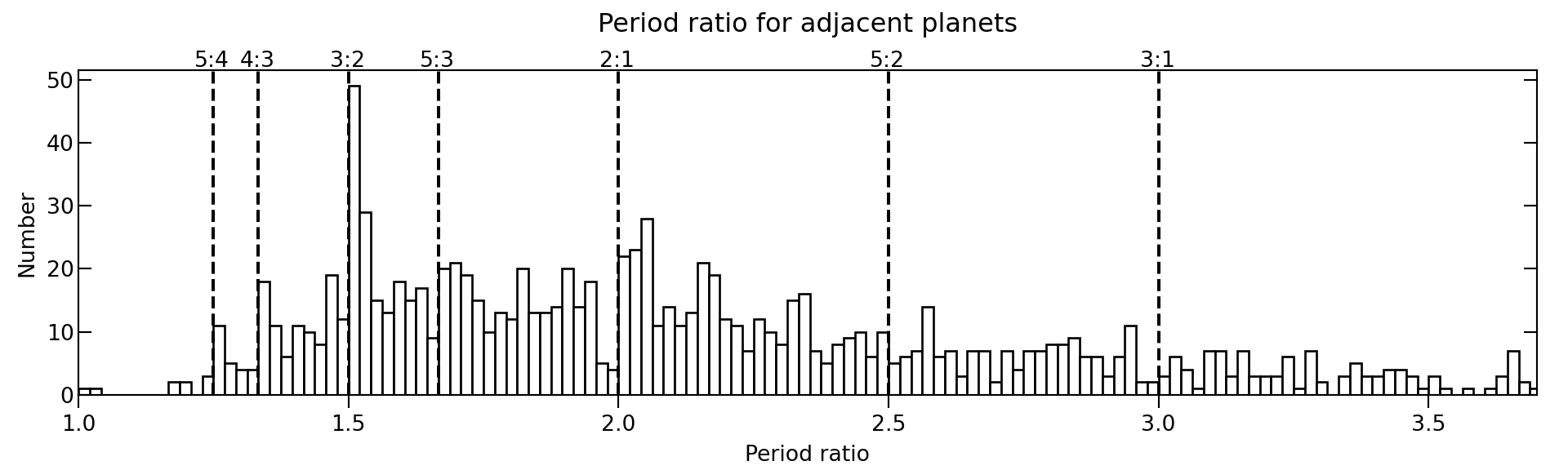}
\caption{Distribution of adjacent-planet period ratios in confirmed multi-planet systems. The data were retrieved from the NASA Exoplanet Archive on June 3, 2026.  }
\label{p-dist}
\end{figure}

One of the most prominent features of compact multi-planet systems is an excess of adjacent planet pairs near low-order integer period ratios, especially the first-order 2:1, 3:2, and 4:3 commensurabilities \citep{FabryckyEtal2014, SteffenHwang2015, HuangOrmel2023, HamerSchlaufman2024, DaiEtal2024}.  The observed distribution of adjacent period ratios (Fig.~\ref{p-dist}) shows clear peaks near these integer period ratios. This clustering is commonly associated with mean-motion resonances (MMRs). Resonant coupling may also involve three planets: in a three-body resonance (3BR), three planets can remain in resonance even when the corresponding two-body resonances are weak or have been disrupted \citep{TamayoEtal2017, Petit2021, GoldbergBatygin2021, HuangOrmel2022, CerioniBeauge2023, YiEtal2025}.

Mean-motion resonances are expected to form during the gas-disk phase {through convergent Type-I migration. Migration plays an important role in the formation of close-in planetary systems \citep{McNeilNelson2010,WangJi2017,RaymondEtal2018,LiuJi2020,WangEtal2021} and, when neighboring planets converge, can assemble them into resonant chains} \citep{GoldreichTremaine1979, LinPapaloizou1979, LeePeale2002, PapaloizouSzuszkiewicz2005, OgiharaKobayashi2013, KajtaziEtal2023, HuangOrmel2023}. A number of systems retain clear examples of such resonance chains, including Kepler-80, Kepler-223, TRAPPIST-1, K2-138, TOI-178, and HD~110067 \citep{MacDonaldEtal2016, MillsEtal2016, GillonEtal2017, ChristiansenEtal2018, LeleuEtal2021TOI, LuqueEtal2023}. These systems provide direct constraints on migration and disk conditions because their resonant architectures preserve information about the formation phase.

Nevertheless, most observed adjacent planet pairs are not in resonance. \citet{DaiEtal2024} found that the occurrence of resonant configurations decreases with system age, suggesting that compact systems form as resonant chains but are subsequently disrupted.
{Early studies of planetary growth, Type-I migration, and disk dispersal already investigated the disruption of resonant chains \citep{OgiharaIda2009,CossouEtal2014}.}
{In particular, \citet{OgiharaIda2009} found that densely packed resonant systems around M dwarfs can undergo orbit crossing and mergers after the gas disk disperses; \citet{CossouEtal2014} found similar post-dispersal instabilities in systems around a solar-mass star.}
{\citet{IzidoroEtal2017} introduced the ``breaking-the-chains'' model, in which migration assembles resonant chains and disk dispersal removes gas damping, allowing dynamical instabilities and giant impacts to transform many of them into non-resonant systems.}

{Subsequent studies have investigated different chain-breaking mechanisms.}
{Possible mechanisms include perturbers on wide orbits \citep{HuangEtal2025,OgiharaKunitomo2026}, disk dispersal and subsequent dynamical instabilities \citep{IzidoroEtal2021,EstevesEtal2022,GoldbergPetit2026}, interactions with planetesimals \citep{GhoshChatterjee2023,LiEtal2026}, instabilities involving higher-order resonances \citep{LiEtal2025,KellerEtal2026}, magnetospheric rebound \citep{LiuEtal2017,LiuOrmel2017,PanEtal2025}, planetary mass loss \citep{MatsumotoOgihara2020,GoldbergBatygin2022,WangLin2023}, and coupling between tidally excited internal oscillations and resonant librations \citep{FarhatChiang2026}. }

Kepler-221 provides an example of a ``break-the-chain'' event: \citet{YiEtal2025} proposed that an original five-planet first-order resonant chain became unstable, that two middle planets merged to form the present planet~d, and that the surviving four planets re-established a non-adjacent 3BR that later expanded through tidal dissipation. A similar scenario has also been proposed for the terrestrial Solar System \citep{HuangEtal2025}. In this "chain-breaking" model, the terrestrial planets started in a resonance chain. Later, perturbations from Jupiter and Saturn broke the resonance chain and led to the merger between Theia and Earth, which resulted in the formation of the Moon.

The recently confirmed planets around Barnard's Star could be another example of a broken resonant chain. Barnard's Star is an old M4 dwarf at a distance of 1.828~pc and is the nearest single star to the Sun \citep{Gizis1997, ByrneEtal2026}. Its age is estimated to be approximately 7--12~Gyr \citep{RiedalEtal2005}. The measured minimum masses, orbital periods, adjacent period ratios, and nearest 3BRs are summarized in Table~\ref{Tab1}.

\begin{table}
\begin{center}
\caption{Dynamical configuration of Barnard's Star planetary system.
The integers $(p,q,r)$ defining the three-body resonance angle are introduced in Eq.~\ref{eq:3BR_define}, and the corresponding $B$-value is defined in Eq.~\ref{eq:B_define}.
The 3BR quantities listed under planets b and c refer to the adjacent triplets d--b--c and b--c--e, respectively.}
\label{Tab1}
\begin{tabular}{ccccc}
\hline
Planet                       & d & b & c & e \\ \hline
$M\sin i$ ($M_\oplus$)       & $0.263\pm 0.024$ & $0.299\pm0.026$ & {$0.335\pm0.030$} & $0.193\pm0.033$ \\
Period (days)                & $2.3402\pm0.0003$ & $3.1542\pm0.0004$ & $4.1244\pm0.0006$ & $6.7392\pm0.0028$ \\
Period Ratio to Inner Planet & \textbackslash{} & 1.3478 & {1.3076} & 1.6340 \\
$(p,q,r)$ of 3BR             &    & (2,3,5) & (5,4,9) & \\
Normalized B-value of 3BR    &    & {$-9.673\times10^{-3}$} & {$-1.443\times10^{-2}$} & \\ \hline
\end{tabular}
\end{center}
\end{table}

Ordered by increasing orbital period, the four planets read d, b, c, and e. The system does not form an obvious chain of adjacent low-order two-body resonances. The innermost pair d--b has a period ratio of 1.3478 and lies close to the 4:3 commensurability, whereas the b--c and c--e pairs are not close to low-order first-order resonances.

In this work, we investigate {four} possible dynamical histories for the Barnard's Star system. {Scenarios~IA and IB test whether tidal expansion or magnetospheric rebound, respectively, can transform an initially resonant four-planet chain into the observed configuration.} Scenarios~{IIA} and {IIB} instead adopt a ``break-the-chain'' interpretation analogous to that proposed for Kepler-221 \citep{YiEtal2025}. We consider an initial five-planet system, denoted d, b, c1, c2, and e from the inside out, in which the resonance chain becomes unstable and planets c1 and c2 merge to form the present-day planet~c. The merger disrupts the intermediate resonant links between c1--c2, while allowing the d--b pair and the non-adjacent b--e pair to remain near resonance.

The difference between Scenarios~{IIA} and {IIB} is the timing and physical origin of the chain-breaking event. In Scenario~{IIA}, the resonant chain survives disk dispersal and expands over Gyr timescales under tidal dissipation; the 3BR breaks only when the system approaches the observed period ratios, triggering the c1--c2 merger. In Scenario~{IIB}, the instability occurs during the disk phase: the disk inner edge migrates outward to the orbit of planet~e, and Lindblad torques acting on the interior planets, particularly c2, disrupt the chain and induce the merger \citep{Ward1997, HuangOrmel2022}.

This paper is organized as follows. In Section~\ref{sect:Model}, we introduce the concept of two-body and three-body resonances, the {four} evolutionary scenarios, and the numerical implementation. Section~\ref{sect:S123} presents the N-body simulations and evaluates the feasibility of each scenario. In Section~\ref{sect:discussion}, we compare the strengths and limitations of the models and discuss their application to other broken-chain systems. Our conclusions are summarized in Section~\ref{sect:conclusion}.

\section{Model}
\label{sect:Model}

\subsection{Two-body and Three-body Resonance}
\label{sect:23R}

Two planets are said to be in a two-body resonance when their orbital period ratio lies close to a ratio of integers and the corresponding resonant angle librates. The two-body resonant angle is defined as:
\begin{equation}
	\phi_{1,2,X} = (j+o)l_2 - jl_1 - o\varpi_X,
	\label{eq:2BR_define}
\end{equation}
where $l_1$ and $l_2$ denote the mean longitudes of the inner and outer planets, respectively, and $\varpi_X$ represents the longitude of periapsis of one of the bodies. The integers $j$ and $o$ are the resonance number and resonance order. For planets trapped in a $(j+o):j$ MMR, the resonant angle $\phi_{1,2,X}$ librates about a fixed value rather than circulating from $0$ to $2\pi$. First-order resonances ($o=1$) are generally more stable than higher-order resonances and more common in planetary systems \citep{MurrayEtal1999}.

The notion of resonance can be extended to three planets by combining two two-body resonant angles. The resulting zeroth-order three-body resonance (3BR) angle is given by \citep{MurrayEtal1999}
\begin{equation}
	\phi_{3\mathrm{BR}} = pl_1 - rl_2 + ql_3,
	\label{eq:3BR_define}
\end{equation}
where $l_i$ denotes the mean longitude of the $i$-th planet; $p$, $q$, and $r$ are integers satisfying $p+q-r=0$. As with the two-body case, three planets locked in a 3BR exhibit libration of this angle. The quantity $p+q-r$ defines the order of the three-body resonance, where zeroth- and first-order 3BRs are typically the most stable \citep{Petit2021}. Observationally, it is easier to detect the time derivative of the 3BR angle, known as the B-value \citep{FabryckyEtal2014, GoldbergBatygin2021}:
\begin{equation}
	B_{3\mathrm{BR}} = p n_1 - r n_2 + q n_3,
	\label{eq:B_define}
\end{equation}
where $n_i$ is the mean motion of the $i$-th planet. The $B$-value quantifies the offset from the three-body resonance, with $B=0$ corresponding to the exact resonance location. For planets participating in a 3BR, the B-value oscillates around zero.

Unlike two-body resonances, zeroth-order three-body resonances do not depend on the longitudes of periapsis and are therefore more readily identifiable from observations. Moreover, 3BRs can persist even when the corresponding two-body resonances have been broken, for instance by tidal orbital expansion \citep{GoldbergBatygin2021, YiEtal2025}. This is because a 3BR corresponds to an independent resonant relation among the three mean motions, and can arise even away from two-body MMRs \citep{Petit2021}.

However, not all resonant chains formed in the disk phase remain intact indefinitely, and observations indicate that 80\% of planets are not in resonance \citep{HuangOrmel2023}. \citet{DaiEtal2024} showed that the fraction of observed resonant chains declines with system age, suggesting that various mechanisms act to disrupt these configurations over time. It is therefore instructive to examine the resonant state of Barnard's Star planetary system in this context.

\subsection{{Scenarios IA and IB: Four-Planet Origins}}
\label{SI}

{We first consider histories in which the system begins with four planets and evolves without a merger. We test two mechanisms: tidal expansion after disk dispersal (Scenario~IA) and magnetospheric rebound during disk dispersal (Scenario~IB).}

Table~\ref{Tab1} summarizes the dynamical architecture of Barnard's Star planetary system \citep{BasantEtal2025}. The innermost planet pair (d and b) lies close to a 4:3 commensurability, whereas the outer adjacent pairs do not reside near any low-order resonances. We have also computed the normalized B-values for the adjacent three-body resonances, defined as $B_\mathrm{nor}=B/\left\langle n\right\rangle$, where $\left\langle n\right\rangle$ is the average of the mean motions of the three planets. These values are listed in Table~\ref{Tab1}. Although the B-values are not as small as those of confirmed 3BR systems (e.g., Kepler-60, Kepler-80, K2-138 \citep{GoldbergBatygin2021}), they are nonetheless relatively close to zero. Stellar tides conserve total angular momentum while dissipating orbital energy and eccentricities, thereby causing the period ratios of planets in a 3BR to expand \citep{CharalambousEtal2018, PapaloizouEtal2018}. If the planets were captured into these 3BRs shortly after disk dispersal, subsequent tidal evolution could have driven the system to its present orbital architecture.

\begin{figure}
\centering
\includegraphics[width=\textwidth]{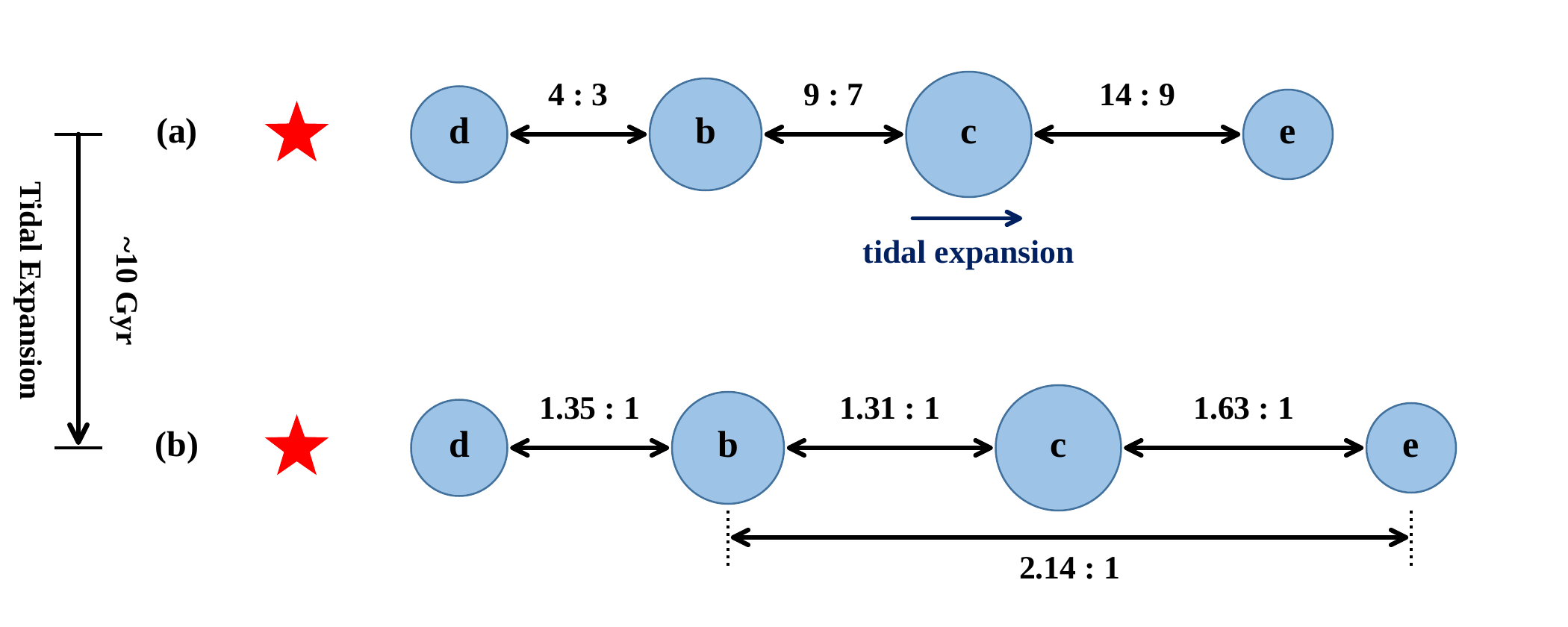}
\caption{Scenario~{IA}: Tidal expansion of an initially resonant four-planet chain toward the observed period ratios. }
\label{Fig-SI}
\end{figure}

We therefore explore the possibility that the current orbital configuration results from tidal expansion of an initially resonant four-planet chain. This evolutionary path is illustrated in Fig.~\ref{Fig-SI} and will be referred to as Scenario~{IA}.

In Scenario~{IA}, the four planets are assumed to have formed a compact MMR chain by the end of the disk phase (panel (a) of Fig.~\ref{Fig-SI}). Tidal dissipation subsequently acts on the planets, causing the resonant chain to expand. {We test whether this expansion can reproduce the observed period ratios.}  A detailed investigation of Scenario~{IA} will be presented in Section~\ref{sect:SI-grand}.

{In Scenario~IB, we begin with a chain of first-order resonances, with adjacent period ratios of 4:3--5:4--3:2 for planets d--b--c--e, and investigate its evolution through magnetospheric rebound \citep{LiuEtal2017,LiuOrmel2017,PanEtal2025}. As the gas accretion rate declines, the stellar magnetic field truncates the disk at progressively larger radii, causing the disk inner edge to move outward. Near this edge, a positive one-sided corotation torque can drive a planet outward with the expanding edge. As the disk becomes less dense, the torque weakens. When the planet can no longer migrate as rapidly as the edge expands, it is left inside the cavity. Planets can decouple from the edge at different times, changing their period ratios and potentially breaking the resonant chain.}

\begin{figure}
\centering
\includegraphics[width=\textwidth]{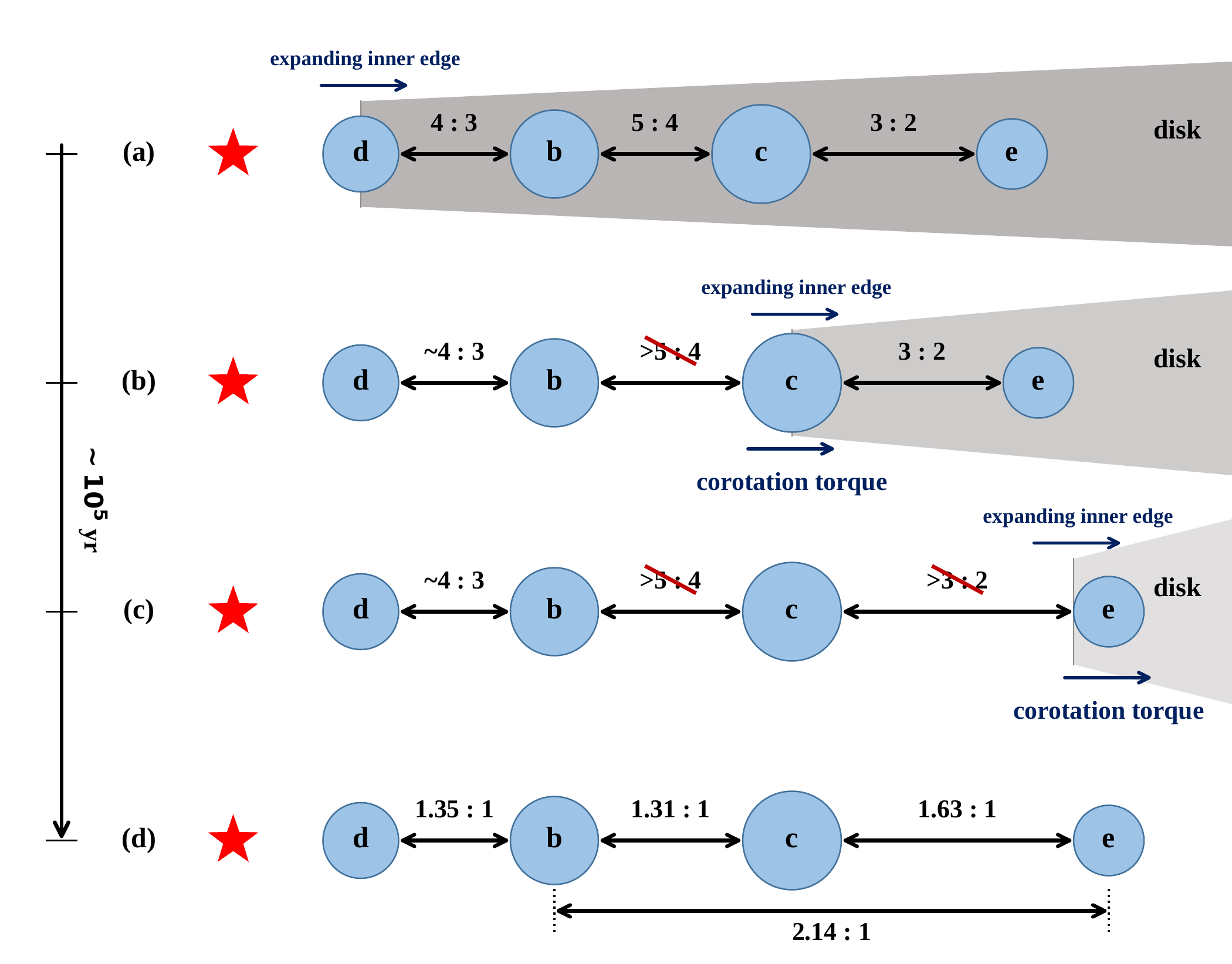}
\caption{{Scenario~IB: Magnetospheric rebound of an initially resonant four-planet chain toward the observed period ratios.}}
\label{Fig-SIB}
\end{figure}

{Figure~\ref{Fig-SIB} illustrates this pathway. Initially, the four planets form a compact resonant chain near the disk inner edge (panel~a). After the edge passes the inner planets, the corotation torque drives planet~c outward, while resonant coupling carries planet~e outward with it (panel~b). Once c is left inside the cavity, the edge can reach e and drive its further outward migration, increasing $P_e/P_c$ (panel~c). Disk dispersal eventually leaves a four-planet system with wider orbital separations (panel~d). In Section~\ref{sect:SIB-grand}, we test whether this mechanism can reproduce all three observed adjacent period ratios simultaneously.}

\begin{figure}
\centering
\includegraphics[width=\textwidth]{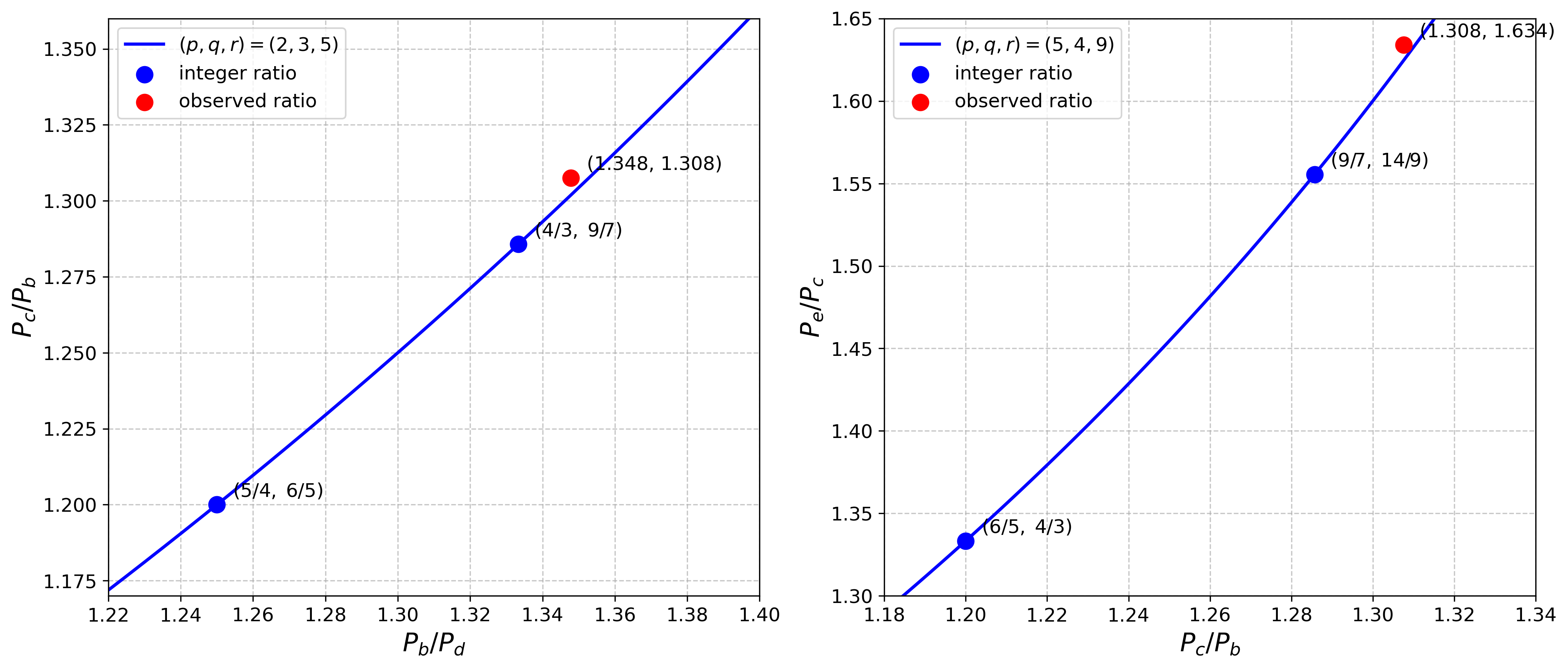}
\caption{Adjacent planet triplets in Barnard's Star system and their closest three-body resonances. The observed period ratios have negligible errors at levels of $10^{-4}$. }
\label{3BR_con}
\end{figure}

\subsection{Scenarios {IIA} \& {IIB}: Five Planets with a Merger Event}
\label{S23}

The architecture of Barnard's Star planets resembles a broken resonant chain: the innermost pair (d, b) lies near a 4:3 resonance and planets b and e are close to a 2:1 commensurability, yet the intermediate adjacent pairs show no clear first-order resonances. This pattern suggests that a planetary merger may have broken part of the original chain while leaving other pairs near their resonant period ratios, analogous to the proposed history of the Kepler-221 system \citep{YiEtal2025} and the early Solar System \citep{HuangEtal2025}.

We therefore hypothesize that the Barnard's Star system initially contained five planets (denoted d, b, c1, c2, and e from inner to outer). These bodies first settled into a stable resonant chain with period ratios near 4:3 (d--b) and 2:1 (b--e). Subsequently, some mechanism destabilized the chain, leading to the merger of planets c1 and c2 to form the present-day planet c. We divide this five-planet model into two sub-scenarios, depending on whether the merger occurred before or after disk dispersal.

\begin{figure}
\centering
\includegraphics[width=\textwidth]{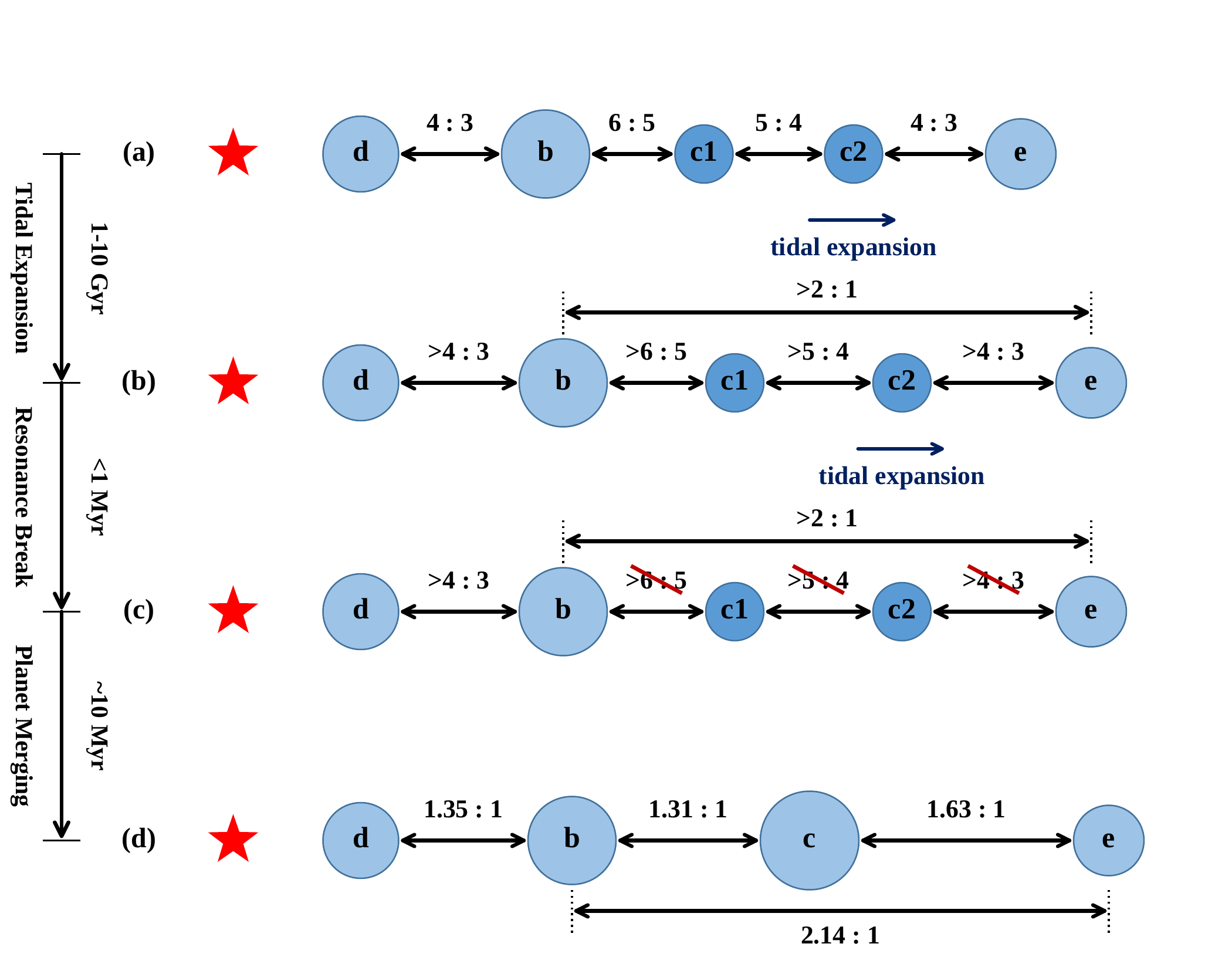}
\caption{Scenario~{IIA}: Tidal expansion disrupts the initial five-planet resonance chain by breaking the adjacent 3BRs. A subsequent merger of c1 and c2 yields the present-day period ratios. }
\label{Fig-SII}
\end{figure}

In Scenario~{IIA} (Fig.~\ref{Fig-SII}), the resonant chain remains intact throughout the disk lifetime, with the planets locked in a sequence of period ratios: 4:3--6:5--5:4--4:3 (panel (a)). Following disk dissipation, Gyr-long tidal evolution expands the 3BR chain, gradually pushing the period ratios toward their current values (panel (b)). The three-body resonance eventually breaks when the period ratios approach the observed state, exciting eccentricities and triggering orbit crossing (panel (c)). Finally, planets c1 and c2 collide and merge, leaving the system with four planets whose period ratios match the observations (panel (d)). Further investigation of Scenario~{IIA} will be provided in Section~\ref{sect:S2-grand}.

\begin{figure}
\centering
\includegraphics[width=\textwidth]{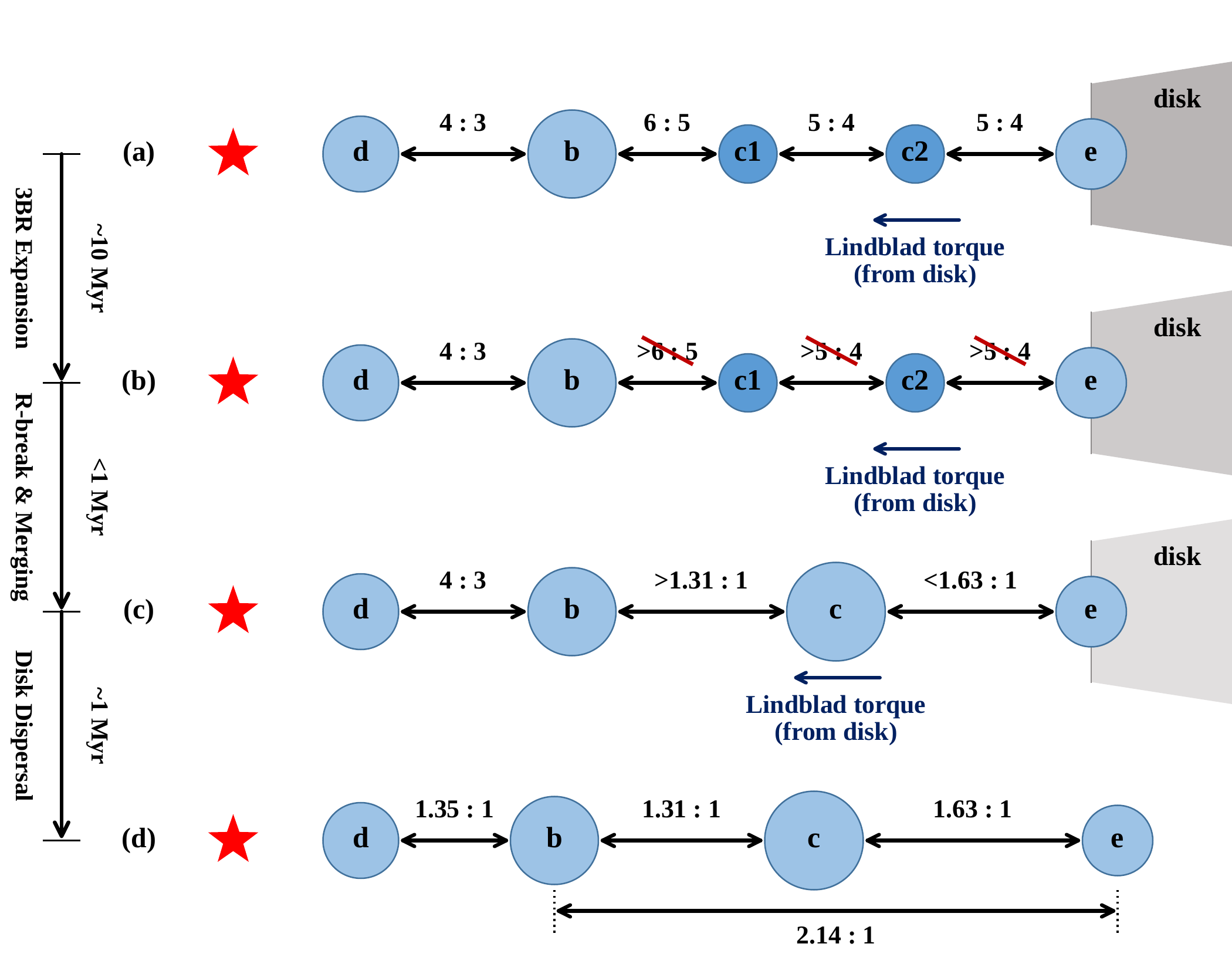}
\caption{Scenario~{IIB}: Lindblad torques expand the inner resonant chain during the disk phase. The resonance between planet c2 and e breaks first, followed by instability in the system and a planet merger. The disk dissipates once the planets approach the observed period ratios. }
\label{Fig-SIII}
\end{figure}

In Scenario~{IIB} (Fig.~\ref{Fig-SIII}), the resonant chain {breaks during the disk phase}. We assume that either the disk inner edge migrated outward or the inner planets were pushed into the cavity, similar to the proposed history of the TRAPPIST-1 system \citep{HuangOrmel2022}. These two pathways are dynamically equivalent in terms of the resulting planetary architecture. The disk inner edge finally resides at the position of planet e. The planets are trapped in a chain of period ratios 4:3--6:5--5:4--5:4 (panel (a)). In this configuration, planet e remains at the disk inner edge \citep{PaardekooperPapaloizou2009, LiuEtal2017}, and the planets interior to the edge experience an inward Lindblad torque \citep{Ward1997, HuangOrmel2022} (see Appendix~\ref{sect:LT}), with planet c2 being particularly affected. This torque drives expansion of the 3BR period ratios and ultimately breaks the chain when certain two-body resonances are crossed \citep{PetitEtal2020i} (panel (b)). Planets c1 and c2 then merge to form planet c (panel (c)). At this stage, the period ratio between c and e remains smaller than the observed value and planet c continues to be pushed inward by the Lindblad torque. The disk finally disperses once the planets have reached orbital periods consistent with the present-day system (panel (d)). Scenario~{IIB} will be discussed in detail later in Section~\ref{sect:S3-grand}.

\subsection{Numerical Implementation}
\label{sect:num}

To investigate the {four} scenarios outlined above, we perform $N$-body simulations using the \texttt{REBOUND} package \citep{ReinLiu2012}. Additional forces are incorporated via the \texttt{REBOUNDx} extension \citep{TamayoEtal2020}. {We use the \texttt{WHFast} integrator \citep{ReinTamayo2015} for simulations without close encounters and the \texttt{TRACE} integrator \citep{LuEtal2024} for simulations involving close encounters.}

For each scenario, we first construct the initial resonant chain through convergent disk-driven migration. The planets are initialized with planet b at the current semi-major axis and the period ratios slightly wider than the target period ratios shown in panel (a) of Figs.~\ref{Fig-SI}, {\ref{Fig-SIB},} \ref{Fig-SII}, and \ref{Fig-SIII}. Migration then drives the planets inward and sequentially captures them into the desired resonances. During this initialization stage, we apply semi-major axis and eccentricity damping using the disk force prescription of \citet{PapaloizouLarwood2000},
\begin{equation}
\mathbf{F}_{\mathrm{disk},i} = -\frac{\mathbf{v}_i}{2t_{a,i}} - \frac{2(\mathbf{v}_i\cdot\mathbf{r}_i)\mathbf{r}_i} {|\mathbf{r}_i|^2 t_{e,i}},
\label{eq:disk_force}
\end{equation}
where $\mathbf{F}_{\mathrm{disk},i}$ is the disk force per unit mass, and $\mathbf{r}_i$ and $\mathbf{v}_i$ are the position and velocity vectors of planet $i$, respectively, and $t_{a,i}$ and $t_{e,i}$ are its semi-major axis and eccentricity damping timescales \citep{HuangOrmel2022}:
\begin{align}
	t_{a,i} &=
	t_{a,\oplus}
	\left(\frac{q_i}{q_\oplus}\right)^{-1},
	\label{eq:ta}\\
	t_{e,i} &= C_e h^2 t_{a,i},
	\label{eq:te}
\end{align}
where $q_i$ and $q_\oplus$ are the planet-to-star mass ratios for planet $i$ and the Earth, $C_e$ is the eccentricity damping prefactor, $h$ is the aspect ratio and $t_{a,\oplus}$ is the nominal Type-I semi-major axis damping timescale for an Earth-mass planet, given by \citep{HuangOrmel2022}:
\begin{equation}
t_{a,\oplus}=1.2\times10^4\left(\frac{\gamma_i}{2}\right)^{-1}\left(\frac{\dot{M}_g}{10^{-10}M_\odot \rm{yr}^{-1}}\right)^{-1}\left(\frac{\alpha}{10^{-3}}\right)\rm{yr},
\label{eq:ta_oplus}
\end{equation}

where $\gamma_i$ is the type-I migration prefactor, $\dot{M}_g$ is the stellar accretion rate and $\alpha$ is the Shakura–Sunyaev viscosity parameter. In this work, the nominal migration timescale is in the range of $t_{a,\oplus}=8\times10^4$--$9\times10^5\,\mathrm{yr}$ (see Table~\ref{Tab2} and Appendix~\ref{sect:LT}).

Once the desired resonant chain has been established, the subsequent evolution differs among the {four} scenarios.

In Scenarios~{IA} and {IIA}, the protoplanetary disk is assumed to disperse after the resonant chain has formed. We therefore remove the disk migration and eccentricity-damping forces in Eq.~\ref{eq:disk_force} and evolve the planets under tidal dissipation. Tidal forces conserve angular momentum while dissipating orbital energy, causing eccentricity damping and a gradual expansion of the resonant chain. The associated eccentricity damping timescale for planet $i$ is given by \citep{PapaloizouEtal2018}
\begin{equation}
	t_e = 7.6\times 10^5 \, Q_{\mathrm{phy}} \frac{m_i}{m_\oplus} \left(\frac{m_\odot}{m_\star}\right)^{1.5} \left(\frac{r_\oplus}{R_i}\right)^5 \left(\frac{a_i}{0.05\,\mathrm{au}}\right)^{6.5} \mathrm{yr},
	\label{eq:t_exp}
\end{equation}
where $m_i$, $R_i$, and $a_i$ are the mass, radius, and semi-major axis of planet $i$, and $m_\star$ is the stellar mass. The parameter $Q_{\mathrm{phy}} = 3Q/(2k_2)$ combines the tidal dissipation function $Q$ and the Love number $k_2$. In our simulations, $Q_{\mathrm{phy}}$ is varied between 10 and 100, comparable to the values inferred for terrestrial planets in the Solar System \citep{GoldreichSoter1966}. To accelerate the initial tidal expansion, we adopt reduced values of $Q_{\mathrm{sim}}/Q_{\mathrm{phy}}$, as listed in Table~\ref{Tab2} \citep{PapaloizouEtal2018}. This acceleration remains valid as long as the system stays locked in the three-body resonance.

{For Scenario~IB, we adopt the magnetospheric rebound prescription of \citet{LiuEtal2017}. During disk dispersal,
\begin{equation}
\dot{M}_g=\dot{M}_{g,0}\exp(-t/\tau_d),\qquad
r_c=r_{c,0}\left(\frac{\dot{M}_g}{\dot{M}_{g,0}}\right)^{-2/7},
\label{eq:IB_disk}
\end{equation}
where $t$ is the time elapsed since the start of the disk dispersal, $\tau_d$ is the disk dispersal timescale, and $r_{c,0}$ is the initial cavity radius.}

{Near the cavity edge, the net one-sided torque on a planet, $\Gamma_{\rm 1s}$, is the sum of the positive corotation torque, $\Gamma_{\rm C,1s}$, and the negative Lindblad torque, $\Gamma_{\rm L,1s}$ \citep{LiuEtal2017}:
\begin{equation}
\begin{aligned}
\Gamma_{\rm 1s} &= \Gamma_{\rm C,1s}+\Gamma_{\rm L,1s},\\
\Gamma_{\rm C,1s} &= 2.46\,ma^2\Omega^2q_d\left(\frac{q}{h^3}\right)^{1/2}\exp(-e/e_f),\\
\Gamma_{\rm L,1s} &= -0.65\,ma^2\Omega^2q_d\frac{q}{h^3}.
\end{aligned}
\label{eq:IB_torque}
\end{equation}
Here $\Gamma_{\rm 1s}>0$ when corotation dominates. For a planet of mass $m$, semi-major axis $a$, and eccentricity $e$, we define $q=m/m_\star$, $q_d=\Sigma a^2/m_\star$, and $\Omega=\sqrt{Gm_\star/a^3}$, where $\Sigma$ is the local gas surface density. The factor $\exp(-e/e_f)$, with $e_f=h/2+0.01$, accounts for eccentricity suppression of the corotation torque \citep{FendykeNelson2014,LiuEtal2017}. The positive net torque adds orbital angular momentum and can carry the planet outward with the expanding disk edge. We also include eccentricity damping following \citet{CresswellNelson2008}.}

For Scenario~{IIB}, the disk remains present after the resonant chain has formed. We assume that the disk inner edge subsequently moves outward or planets migrate into the disk cavity. As a result, planet~e is trapped at the disk edge, while the inner four planets reside inside the disk cavity. Planets inside the cavity can still interact with the exterior disk through outer Lindblad resonances that remain within the gas disk \citep{Ward1997,HuangOrmel2022}. We parameterize the resulting inward migration torque as
\begin{equation}
	\Gamma_L = \;\sum_{m\,\in\,\mathrm{disk}} f_m\Gamma_{a,i}
	\label{eq:LT_main}
\end{equation}
where $\Gamma_{a,i}$ is the migration torque acting on planet $i$, related to its semi-major axis damping timescale by
\begin{equation}
    \Gamma_{a,i} = -\frac{L_i}{2t_{a,i}},
    \label{eq:LT_Gamma}
\end{equation}
where $L_i$ is the orbital angular momentum and the semi-major axis damping timescale $t_{a,i}$ is scaled from the nominal migration timescale $t_{a,\oplus}$. $f_m$ is the normalized contribution from the $m$-th outer Lindblad resonance. The sum includes only resonances whose locations lie within the gas disk (see Appendix~\ref{sect:LT}). As a planet moves away from the disk inner edge and deeper into the cavity, fewer Lindblad torque modes remain within the gas disk, causing the total one-sided torque to decrease.

As indicated in Table~\ref{Tab1}, the planetary masses are degenerate with the orbital inclination $i$, because radial-velocity observations constrain only $M\sin i$. The inclinations adopted for each scenario are listed in Table~\ref{Tab2}. Planetary radii are estimated using the mass--radius relation in \citet{ChenKipping2017}. The full range of parameter values explored in our $N$-body simulations is summarized in Table~\ref{Tab2}.

During the evolution in each scenario, the semi-major axis of planet~b can shift slightly as a result of tidal evolution or disk torques. We therefore compare the final configurations with the observed system using the planetary period ratios, rather than requiring an exact match in semi-major axis.

\begin{table}
\centering
\caption{
Parameter values and ranges adopted in the simulations of the {four} scenarios.
{Migration timescales $t_{a,\oplus}$ are in units of $10^4$\,yr, and accretion rates $\dot M_{g,0}$ are in units of $10^{-10}\,M_\odot\,\mathrm{yr}^{-1}$.}
A dash indicates that the parameter is not applicable to that scenario.
}
\label{Tab2}
\small
\setlength{\tabcolsep}{3pt}
\renewcommand{\arraystretch}{1.35}
\begin{tabular*}{\textwidth}{@{\extracolsep{\fill}}ccccccccc@{}}
\hline
Scenario
& $i$
& $Q_{\mathrm{phy}}$
& $Q_{\mathrm{sim}}/Q_{\mathrm{phy}}$
& $m_{\mathrm{c1}}/m_{\mathrm{c}}$
& $t_{a,\oplus}$
& {$\dot M_{g,0}$}
& {$h$}
& Figures \\
\hline

{IA}
& $45^\circ,\ 90^\circ$
& $[10,100]$
& $10^{-3}$
& --
& --
& {--}
& {--}
& \ref{Fig-SI_P}--\ref{Fig-SI_B} \\

{IB}
& {$45^\circ,\ 90^\circ$}
& {--}
& {--}
& {--}
& {--}
& {$[1,2]$}
& {$[0.025,0.030]$}
& {\ref{Fig-SIB_evo}--\ref{Fig-SIB_dist}} \\

{IIA}
& $45^\circ,\ 60^\circ,\ 90^\circ$
& $[10,60]$
& $[3\times10^{-4},10^{-3}]$
& $[0.75,0.85]$
& --
& {--}
& {--}
& \ref{Fig-SII_semi}--\ref{Fig-SII_dist} \\

{IIB}
& $45^\circ,\ 90^\circ$
& --
& --
& $[0.6,0.8]$
& $[8,90]$
& {--}
& {--}
& \ref{Fig-SIII_semi}--\ref{Fig-SIII_ta}  \\
\hline
\end{tabular*}
\end{table}

\section{Evolution of Barnard's Star System}
\label{sect:S123}

\subsection{Scenario {IA}: Tidal Expansion with Four Planets}
\label{sect:SI-grand}
In this section, we examine whether the current orbital architecture of Barnard's Star planets can be explained by pure three‑body resonance (3BR) expansion of an initial four‑planet chain (Scenario~{IA}, Fig.~\ref{Fig-SI}).

We initialize the planets with period ratios close to those shown in panel (a) of Fig.~\ref{Fig-SI}. In this configuration, the inner three planets are near the $(2,3,5)$ 3BR and the outer three are near the $(5,4,9)$ 3BR, where the integers $(p,q,r)$ follow the definition in Eq.~(\ref{eq:3BR_define}). Tidal damping (Sect.~\ref{sect:num}) is then applied after disk dispersal, causing the planets to expand along the respective 3BR lines (see Fig.~\ref{3BR_con}). In order to reproduce the observed period ratios, the expansion must satisfy the following criteria:
\begin{enumerate}
    \item The expansion must be completed within the age of Barnard's Star, which is estimated to be $7$–$12$ Gyr \citep{RiedalEtal2005}.
    \item The 3BR expansion should successfully reach the observed period ratios without breaking.
    \item The libration amplitude of the $B$-value (Eq.~\ref{eq:B_define}) must be large enough to cover the measured normalized values listed in Table~\ref{Tab1}.
\end{enumerate}

To meet the first criterion, we adopt $Q_{\mathrm{phy}} = 20$–$30$, which allows the planets to approach the observed period ratios within $10$\,Gyr. To speed up the numerical integration, we use a reduced tidal parameter $Q_{\mathrm{sim}} = 10^{-3}\,Q_{\mathrm{phy}}$ during the early expansion phase. We performed 100 independent simulations, varying $Q_{\mathrm{phy}}$, the planetary masses (through the inclination $i$; Table~\ref{Tab2}), and the initial period ratios. However, criteria 2 and 3 were not satisfied: (i) in 90\% of the simulations, the 3BR broke before the planets reached the observed ratios, and (ii) in the remaining 10\%, the libration amplitude of the $B$-value remained too small to cover the observed offset from exact resonance.

\begin{figure}
\centering
\includegraphics[width=\textwidth]{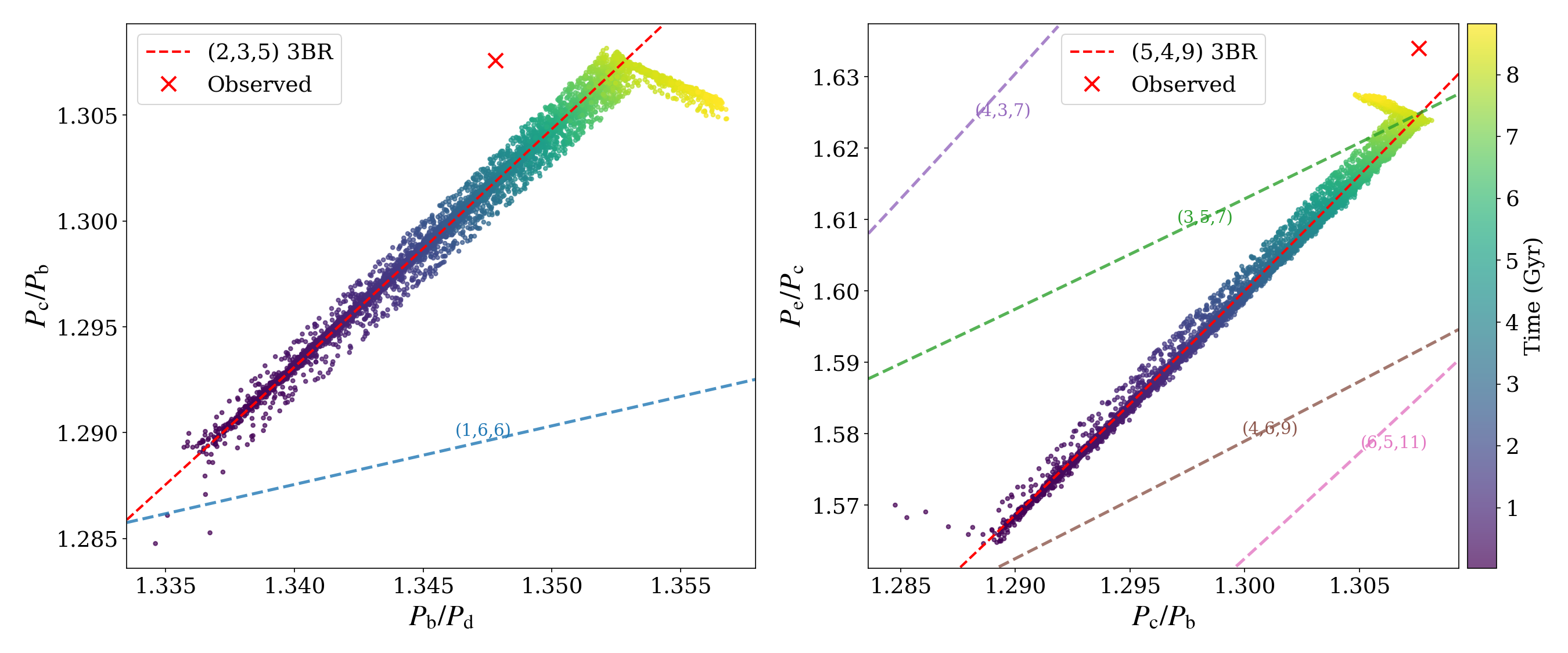}
\caption{Evolution of the period ratios in one example simulation of Scenario~{IA} ($i=45^\circ$, $Q_{\mathrm{phy}}=20$). Left panel: period ratios of planets~d, b, and c; right panel: period ratios of b, c, and e. The red dashed lines represent three-body resonances ($(p,q,r)=(2,3,5)$ for d--b--c and $(5,4,9)$ for b--c--e). Other dashed lines show other close three-body resonances. The planets expand along the resonant lines until $\sim 8$\,Gyr, after which the resonance breaks.}
\label{Fig-SI_P}
\end{figure}

\begin{figure}
\centering
\includegraphics[width=\textwidth]{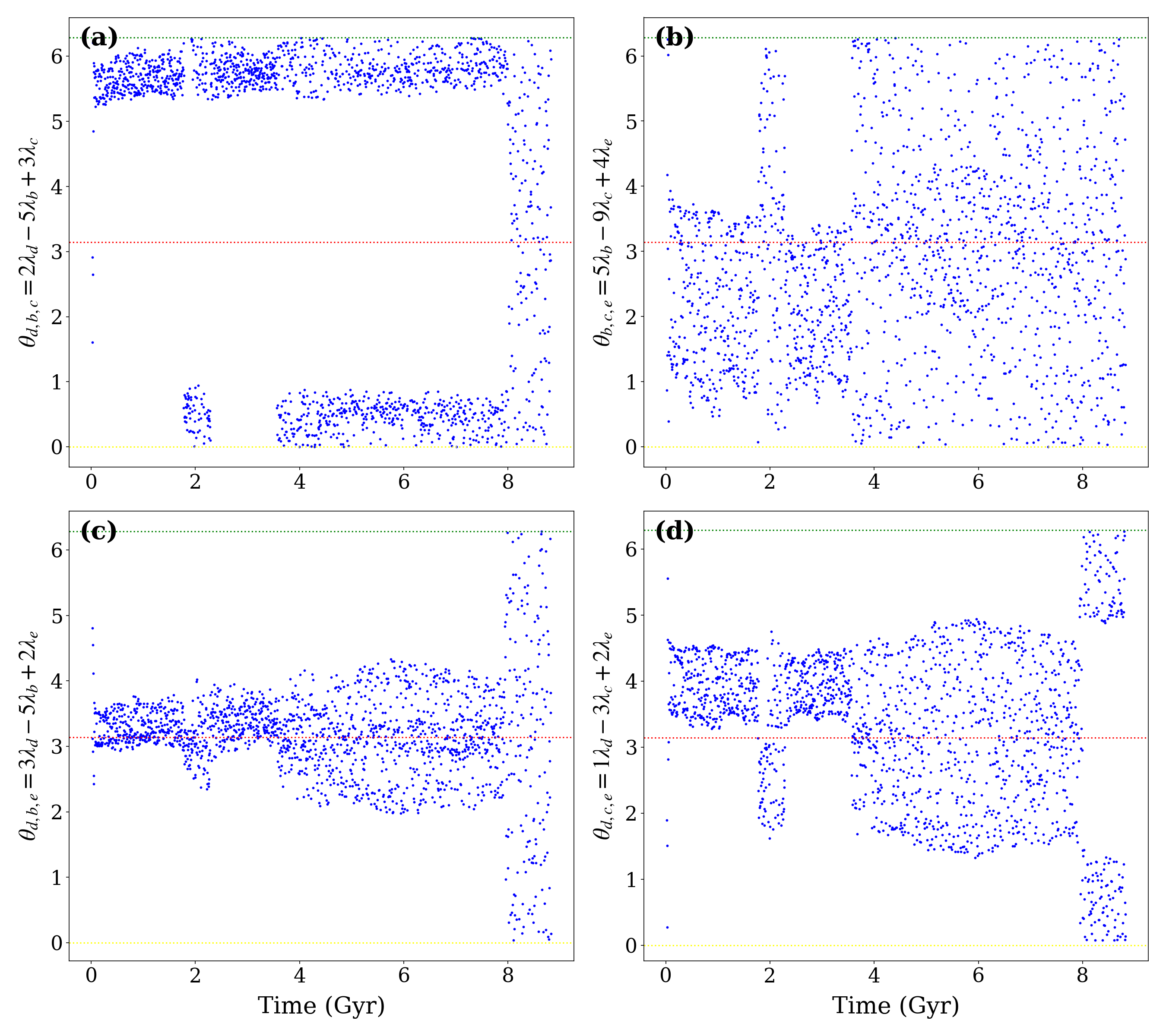}
\caption{Evolution of the three-body resonance angles for the same simulation as Fig~\ref{Fig-SI_P}. The 3BR angles for the adjacent triplets (d, b, c) and (b, c, e) are shown, together with the non‑adjacent combinations (d, b, e) and (d, c, e). The adjacent (d, b, c) and non‑adjacent 3BRs librate and drive the resonant expansion until $\sim 8$\,Gyr.}
\label{Fig-SI_3BR}
\end{figure}

\begin{figure}
\centering
\includegraphics[width=\textwidth]{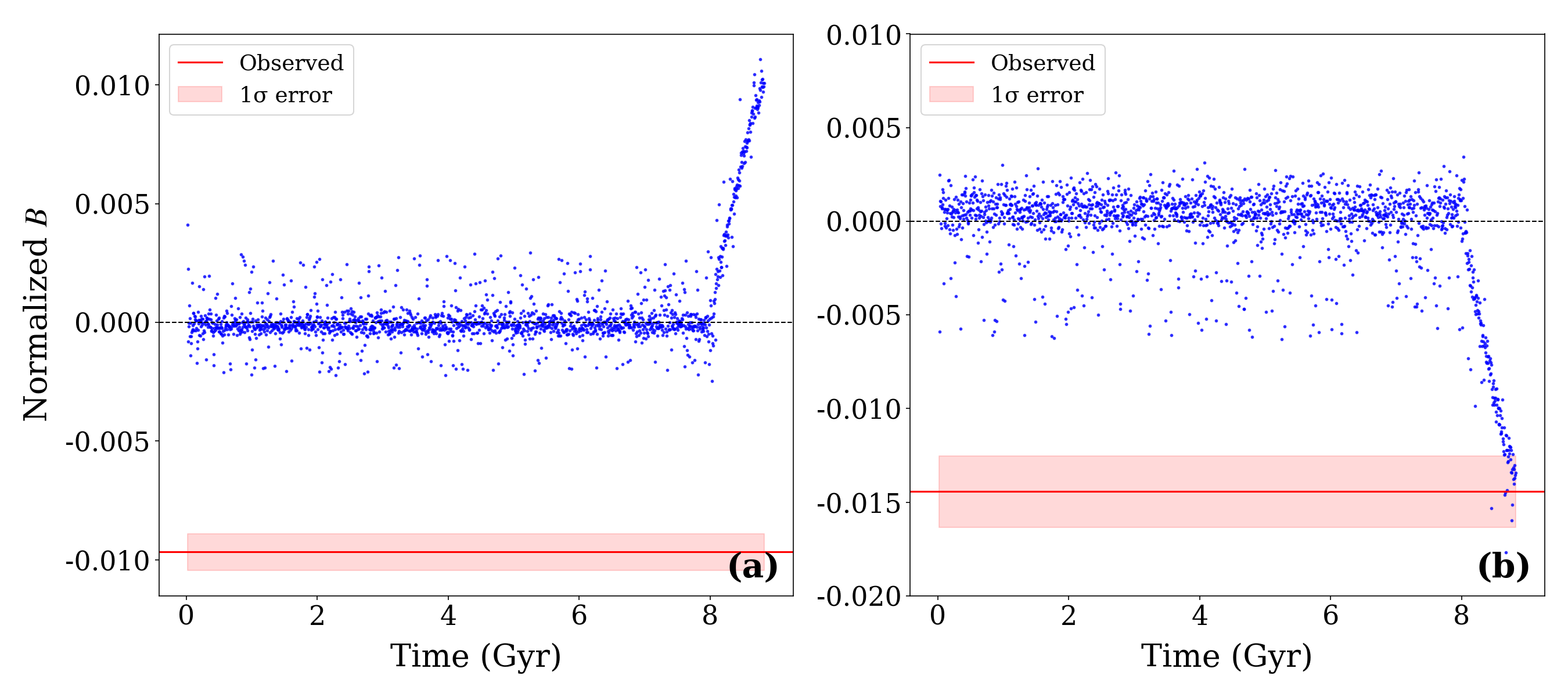}
\caption{Evolution of the normalized $B$-value (Eq.~\ref{eq:B_define}) for the same simulation as Fig.~\ref{Fig-SI_P} for the adjacent triplets (d, b, c) and (b, c, e). The shaded bands indicate the $1\sigma$ observed ranges from Table~\ref{Tab1}. The libration amplitude never reaches the observed offset, even before the resonance breaks.}
\label{Fig-SI_B}
\end{figure}

Figs~\ref{Fig-SI_P}, \ref{Fig-SI_3BR}, and \ref{Fig-SI_B} illustrate the evolution for one representative case with $i = 45^\circ$ and $Q_{\mathrm{phy}} = 20$. Fig.~\ref{Fig-SI_P} shows the period ratio evolution; the planets expand along the 3BR lines (red dashed) until $\sim 8$\,Gyr, at which point the resonance breaks because the outer three planets are perturbed by the first-order $(3,5,7)$ 3BR. After the break, tidal damping drives the period ratio between planets~d and~b further upward, failing to reproduce the observed configuration.

Fig.~\ref{Fig-SI_3BR} displays the evolution of the three-body resonance angles. Notably, the adjacent 3BR between b, c, and e never librates; instead, the lower‑order and stronger non‑adjacent 3BRs (d, b, e and d, c, e) librate and sustain the resonant expansion for the first $8$\,Gyr before breaking. The non-adjacent 3BR between planets d, c and e even stays in resonance after $8$\,Gyr.

Fig.~\ref{Fig-SI_B} tracks the $B$-value evolution, with the shaded region indicating the observed range. Even before the 3BR breaks, the libration amplitude is far too small to cover the observed offset from zero.

From these analyses we conclude that it is unlikely that Barnard's Star planets originated as a four‑planet system that reached the observed period ratios solely through tidal expansion of a pure 3BR chain. {We next examine whether magnetospheric rebound during disk dispersal can reproduce the observed configuration in Scenario IB.}

\subsection{{Scenario IB: Magnetospheric Rebound}}
\label{sect:SIB-grand}

{We now investigate Scenario~IB, in which four planets initially form a compact first-order resonant chain. As illustrated in Fig.~\ref{Fig-SIB}, magnetospheric rebound drives the outer two planets outward as the disk inner edge expands, potentially bringing their period ratios into agreement with the observed values.}

{We carried out 18 $N$-body simulations with the parameter combinations listed in Table~\ref{Tab2}. We explore $h=0.025$--$0.030$, starting from the value adopted by \citet{LiuEtal2017}, and use $\tau_d=3\times10^4$\,yr and $\dot{M}_{g,0}=(1$--$2)\times10^{-10}\,M_\odot\,\mathrm{yr}^{-1}$, selected from a broader parameter search for configurations closest to the observed period ratios.}
{After removing the gas forces, we evolve each system for another $10^4$\,yr and average the period ratios over the final 200\,yr.}
{We quantify the agreement with the observed configuration using the sum of squared fractional period-ratio differences,
\begin{equation}
 dP=\sum_{j=1}^{3}\left[\frac{\left(P_{j+1}/P_j\right)_{\mathrm{sim}}-\left(P_{j+1}/P_j\right)_{\mathrm{obs}}}{\left(P_{j+1}/P_j\right)_{\mathrm{obs}}}\right]^2,
 \label{eq:SIII-dP}
\end{equation}
where the planets are ordered as d, b, c, and e. A smaller $dP$ indicates a closer match.}

\begin{figure}
\centering
\includegraphics[width=\textwidth]{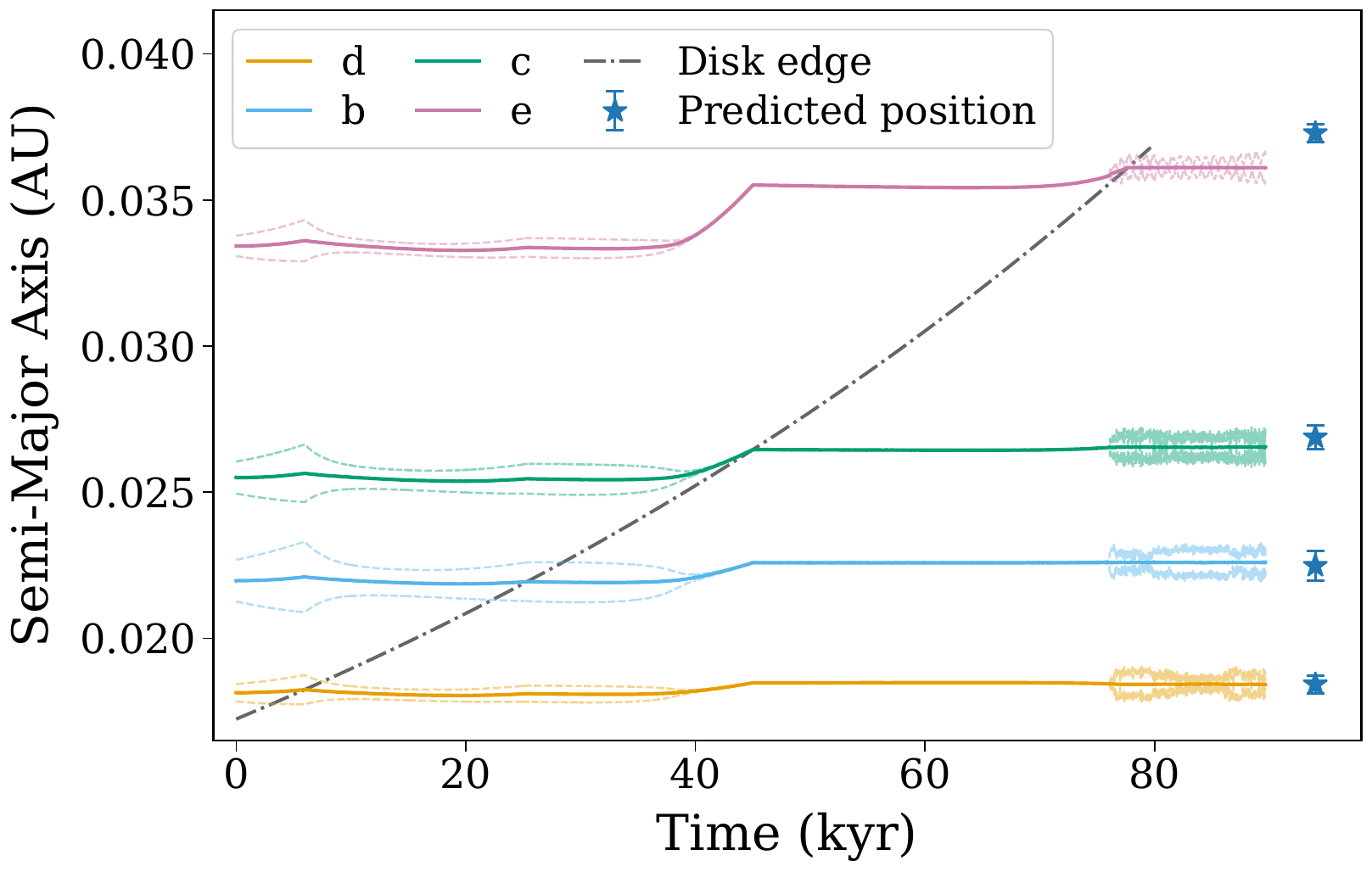}
\caption{{Evolution of the semi-major axes (solid lines), pericenter and apocenter (dashed lines) in the closest-matching Scenario~IB simulation. The dash-dotted line marks the disk inner edge.}}
\label{Fig-SIB_evo}
\end{figure}

{Figure~\ref{Fig-SIB_evo} shows the closest match, with the smallest $dP$ among the 18 simulations ($i=45^\circ$, $\dot M_{g,0}=1.5\times10^{-10}\,M_\odot\,\mathrm{yr}^{-1}$, and $h=0.0275$).}
{Its final outermost period ratio is $P_e/P_c\simeq1.587$, which is lower than the observed value of 1.634.}
{This mismatch arises from the mass dependence of corotation-driven migration: lower-mass planets migrate outward more slowly (Eq.~\ref{eq:IB_torque}), so planet~e, the least massive planet, cannot keep pace with the expanding disk edge as the gas density declines.}

\begin{figure}
\centering
\includegraphics[width=\textwidth]{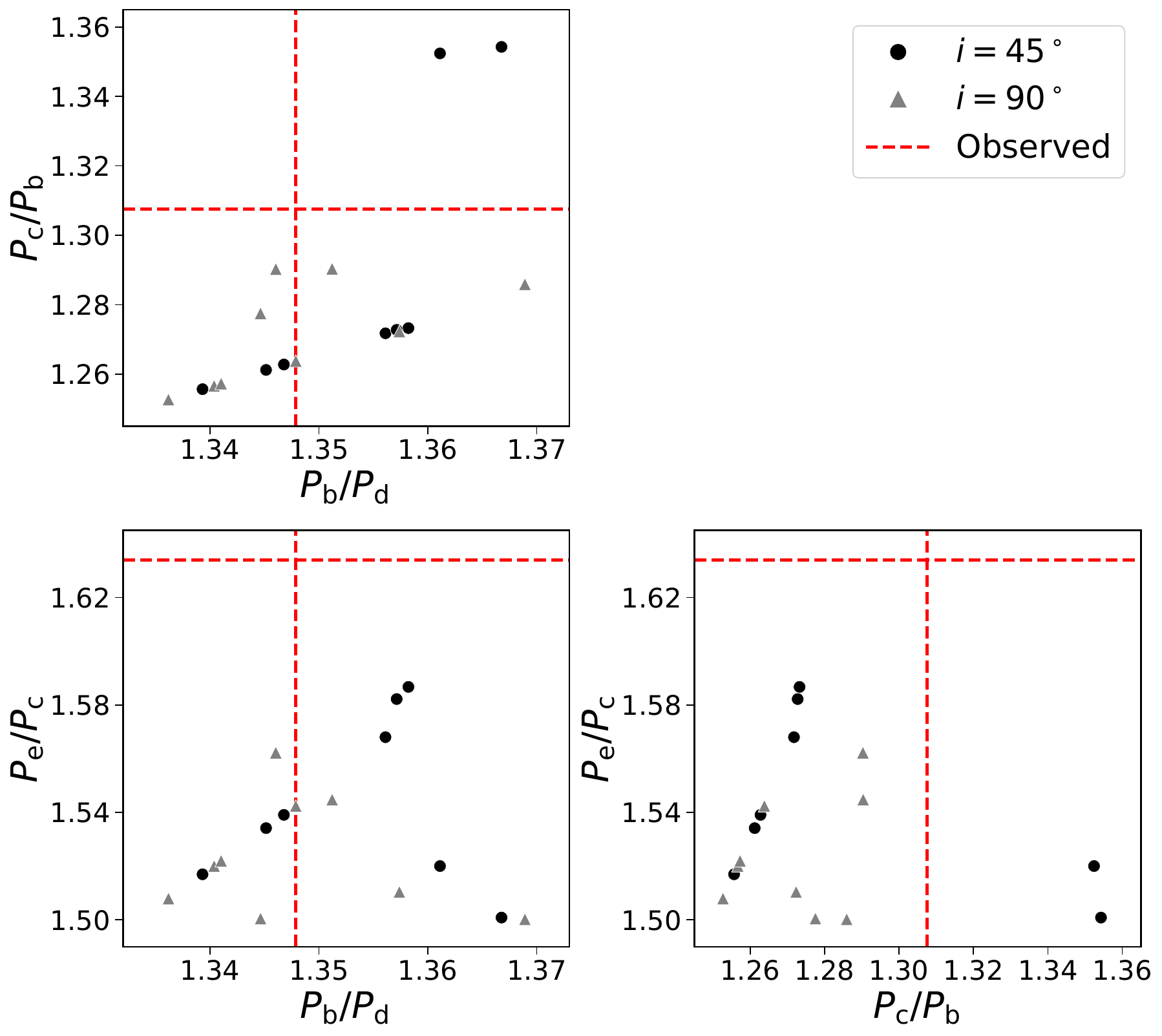}
\caption{{Final adjacent period ratios for Scenario~IB. Black circles and gray triangles denote $i=45^\circ$ and $90^\circ$, respectively; red dashed lines mark the observed values. One simulation, with $P_c/P_b=1.637$, lies outside the displayed ranges.}}
\label{Fig-SIB_dist}
\end{figure}

{Figure~\ref{Fig-SIB_dist} shows the final adjacent period ratios for the full sample. All 18 simulations leave the outermost pair too closely spaced, with $P_e/P_c\leq1.587$. Slowing the disk-edge expansion after it passes planet~c can bring the outermost ratio closer to the observed value, but this requires additional tuning of the dispersal history. Thus, we do not include such runs here. We therefore find that magnetospheric rebound does not reproduce the observed configuration within the tested parameter grid.}

\subsection{Scenario {IIA}: Tidal Expansion ``Break-the-chain''}
\label{sect:S2-grand}

In this section, we investigate Scenario {IIA} (shown in Fig.~\ref{Fig-SII}). In this scenario, the system originally contained five planets in resonance (namely d, b, c1, c2, and e), and a subsequent merger of c1 and c2, triggered by tidal expansion, produced the present‑day architecture.

We initialize the five planets in a first‑order resonance chain of period ratios 4:3 – 6:5 – 5:4 – 4:3 (panel (a) of Fig.~\ref{Fig-SII}). In this configuration, all adjacent planet triplets are simultaneously locked in three‑body resonances. Tidal damping then expands the planets along the 3BR, driving the period ratios of planets~d, b and e toward the observed ratios. For the orbital configuration to match the observations, the following three requirements must be fulfilled:

\begin{enumerate}
    \item The three‑body resonances between the planets break when their period ratios approach the observed values.
    \item Orbital eccentricities build up after the 3BR break, leading to orbit crossing and a successful merger of planets~c1 and~c2.
    \item The system remains dynamically stable thereafter, with no additional mergers.
\end{enumerate}

We test this scenario with $N$-body simulations. We performed 75 independent runs, varying the planetary masses (through the inclination $i$), the tidal quality factor $Q_{\mathrm{phy}}$, and the mass ratio $m_{\mathrm{c1}}/m_{\mathrm{c}}$ (Table~\ref{Tab2}). To accelerate the initial expansion, we use $Q_{\mathrm{sim}}/Q_{\mathrm{phy}}\approx3\times10^{-4}$--$10^{-3}$. A simulation was stopped if a different pair of planets merged or if multiple mergers occurred. When planets~c1 and~c2 successfully merged, we continued the integration for an additional $10$\,Myr to verify the long‑term stability of the final four‑planet system.

\begin{figure}
\centering
\includegraphics[width=\textwidth]{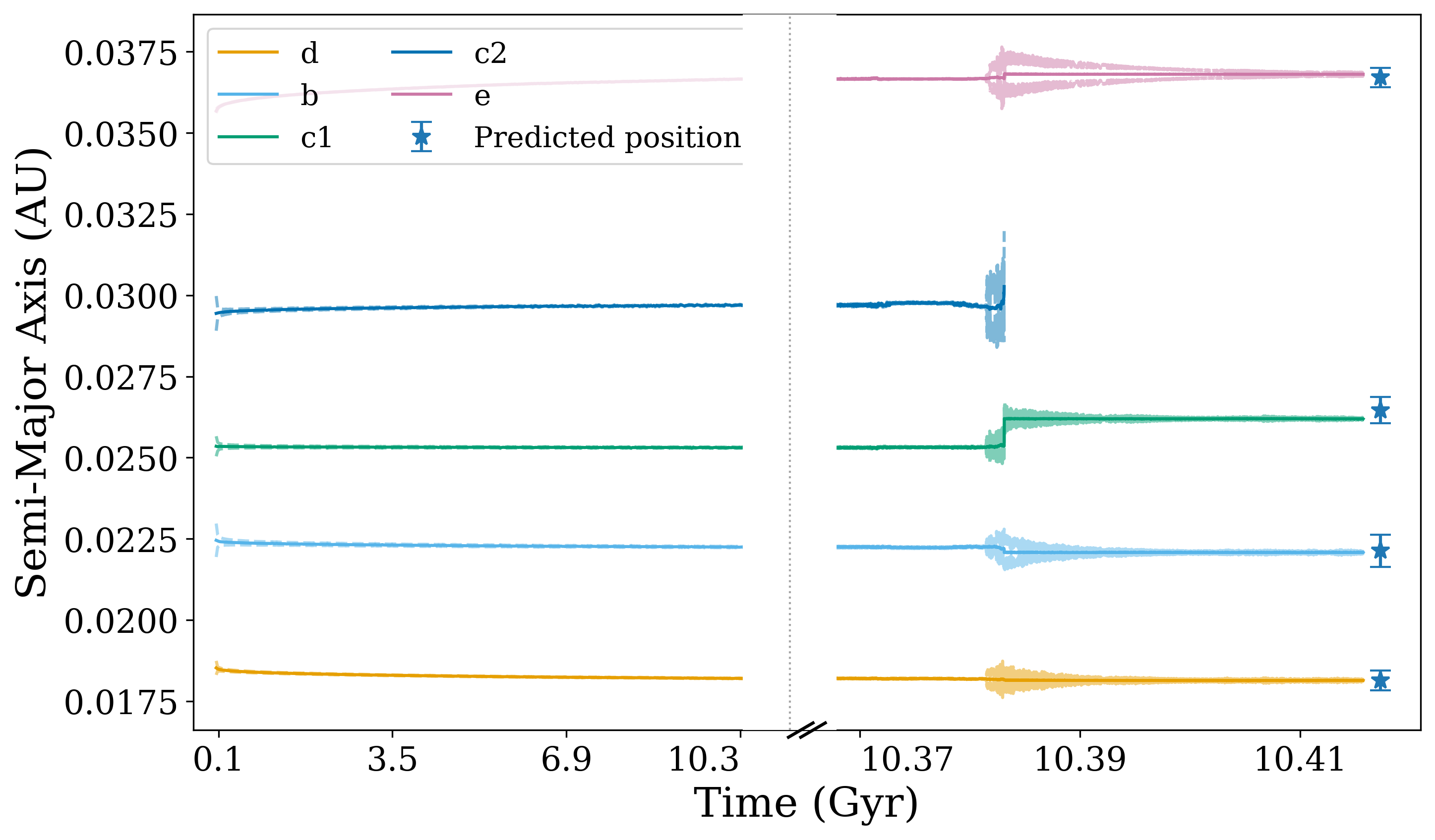}
\caption{Evolution of the semi‑major axes (solid lines), pericenter and apocenter (dashed lines) of the planets in one successful example in Scenario {IIA} ($i=45^\circ$, $m_{\mathrm{c1}}/m_{\mathrm{c}}=0.825$). The left panel shows the tidal expansion phase and the right panel shows the merging phase.  }
\label{Fig-SII_semi}
\end{figure}

Fig.~\ref{Fig-SII_semi} presents an example simulation in which c1 and c2 merge to form a planet~c that is close to the observed orbital period, and the final system remains stable. The left panel shows the expansion phase, during which the period ratios increase while the planets stay trapped in the 3BR. The right panel zooms into the merging phase, where the 3BR has already broken. In this phase, planets~c1 and~c2 collide, and the resulting four‑planet system remains stable without further merging events.

\begin{figure}
\centering
\includegraphics[width=\textwidth]{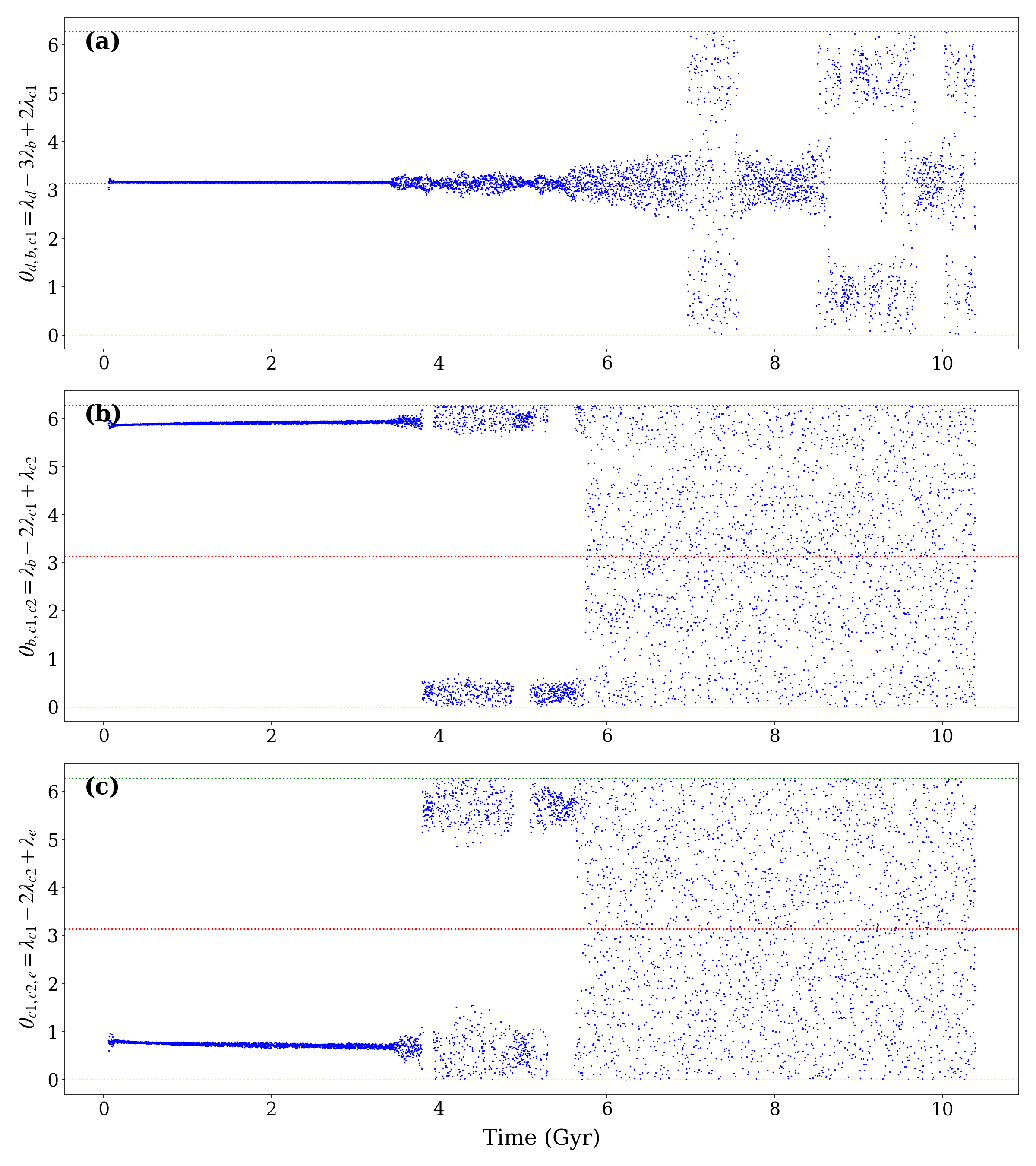}
\caption{Evolution of the adjacent three‑body resonance angles for the same simulation as Fig.~\ref{Fig-SII_semi} in Scenario {IIA}. The angles evolve from libration to a switching state (see Appendix~\ref{sect:3BR_mode}) and finally evolve to circulation, which triggers the merger event. }
\label{Fig-SII_R}
\end{figure}

Fig.~\ref{Fig-SII_R} shows the evolution of the adjacent three‑body resonance angles in the same simulation. The angles clearly librate for the first $5.5$\,Gyr. From $5.5$\,Gyr to $10$\,Gyr, the resonance appears to be broken, but in reality the angles remain librating, with rapidly switching libration centers that mimic circulation over longer intervals. This behavior is described in detail in Appendix~\ref{sect:3BR_mode}. Such a switching state allows the period ratios to continue expanding along the 3BR until the resonance finally breaks completely at the end of the simulation (around ${\sim}10$\,Gyr).

\begin{figure}
\centering
\includegraphics[width=\textwidth]{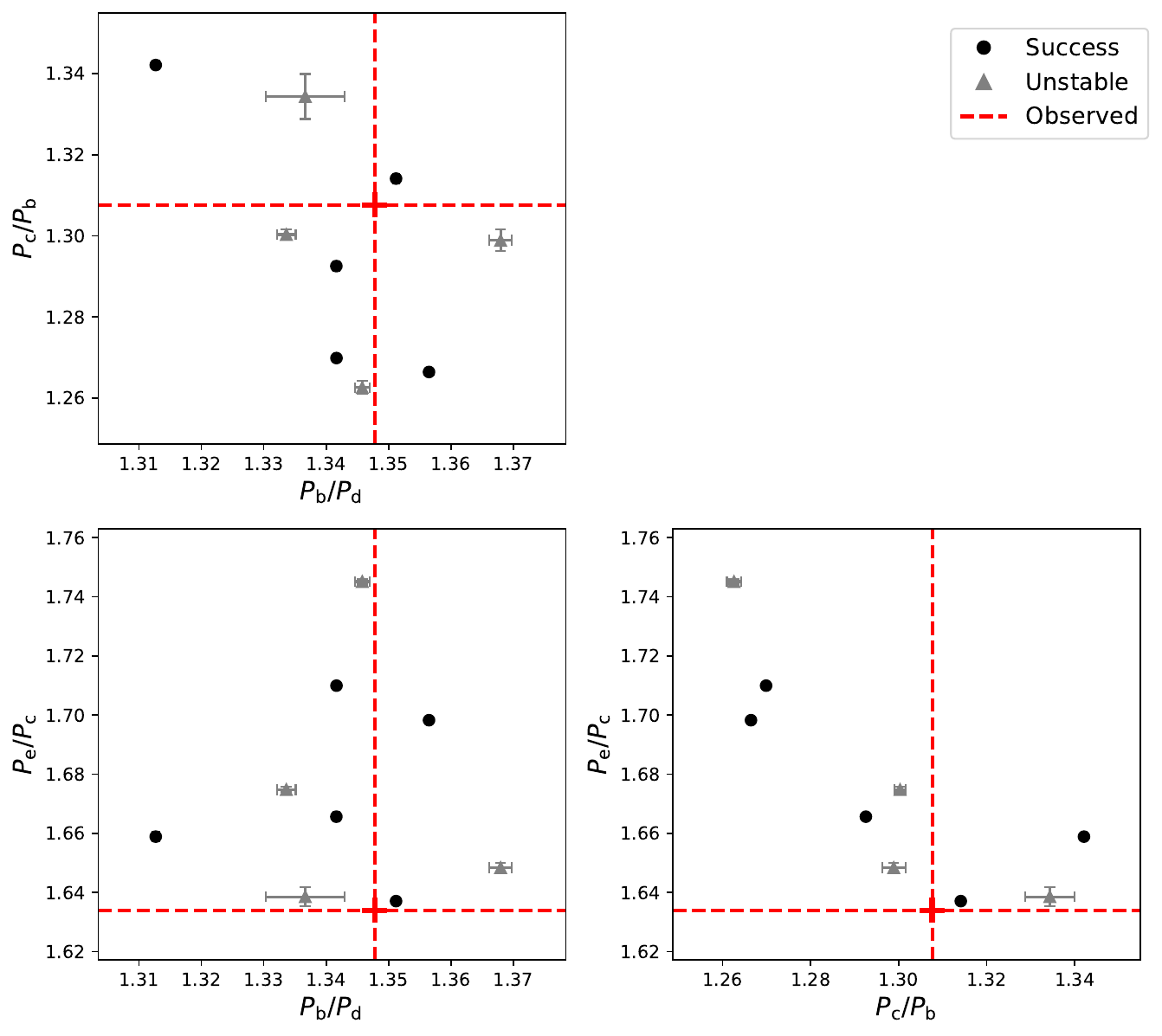}
\caption{Distribution of adjacent planet period ratios measured $0.1$\,Myr after the c1 \& c2 merger events for Scenario {IIA}. Black circles denote long-term stable simulations that remained stable for $10$\,Myr after the merger; gray triangles indicate unstable simulations where a subsequent merger occurred. The error bars show the libration amplitude of the period ratios. }
\label{Fig-SII_dist}
\end{figure}

Of the 75 simulations, 30 experienced a merger event. In 9 of these, the merger involved only c1 and c2, and 5 of those 9 remained stable for the subsequent $10$\,Myr. Fig.~\ref{Fig-SII_dist} displays the distribution of adjacent period ratios for the 9 successful mergers, compared with the observed values. The ratios generally agree with the observations, although the outermost period ratio ($P_e/P_c$) tends to be slightly larger than the measured value. This discrepancy may arise because the mass ratio $m_{\mathrm{c1}}/m_{\mathrm{c}}$ is lower than the tested value in reality.

In summary, Scenario~{IIA} provides a plausible evolutionary path where tidal expansion after disk dispersal can naturally “break the chain” and trigger the merger of c1 and c2, ultimately yielding the current four‑planet system. The final ratios depend on parameters including the planetary masses and the mass ratio $m_{\mathrm{c1}}/m_{\mathrm{c}}$, but in general agree with the observed configuration in the Barnard's Star system.

\subsection{Scenario {IIB}: Lindblad Torque ``Break-the-chain''}
\label{sect:S3-grand}
In Scenario~{IIA}, the ``break-the-chain'' event occurs after disk dispersal. Here we investigate an alternative (Scenario~{IIB}) in which the resonant chain breaks already during the disk phase, as illustrated in Fig.~\ref{Fig-SIII}.

The five planets are initialized in a first‑order resonance chain of period ratios 4:3--6:5--5:4--5:4 (panel (a) of Fig.~\ref{Fig-SIII}). Compared to Scenario~{IIA}, the outermost pair is initialized in a more compact 5:4 resonance rather than 4:3. If it instead started from 4:3, the strong Lindblad-torque-driven expansion of the c2--e pair would move the planets too far apart, leaving the final c--e period ratio significantly larger than the observed value.

\begin{figure}
\centering
\includegraphics[width=\textwidth]{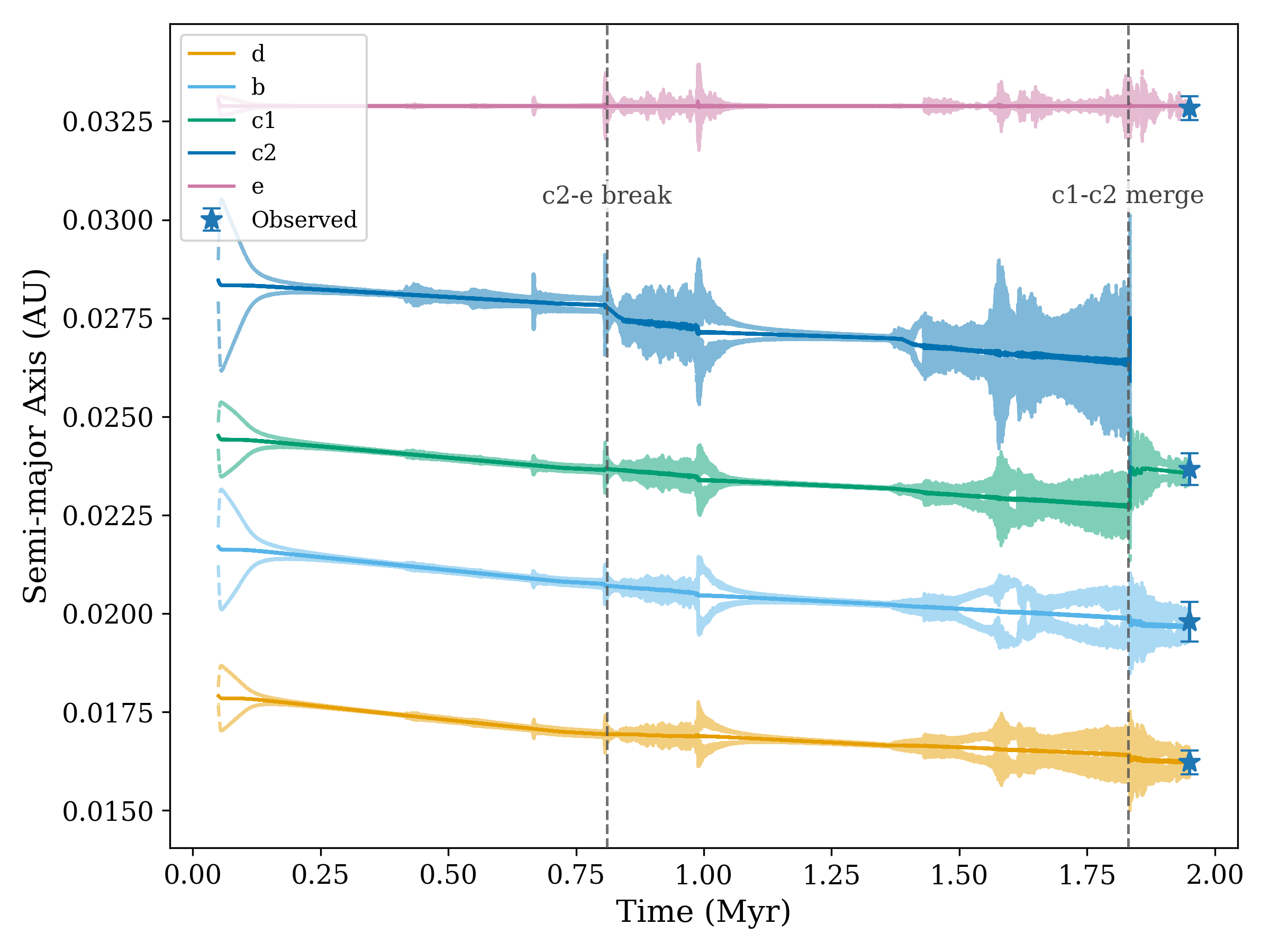}
\caption{Evolution of the semi‑major axes (solid lines), pericenter and apocenter (dashed lines) in a successful Scenario~{IIB} simulation ($i=45^\circ$, $m_{\mathrm{c1}}/m_{\mathrm{c}}=0.8$, $t_{a,\oplus}=2.8\times 10^5\,\mathrm{yr}$). }
\label{Fig-SIII_semi}
\end{figure}

After the resonance chain is established, the planets enter the disk cavity either because the disk inner edge migrates outward or because the planets migrate inward, similar to the scenario invoked for the TRAPPIST‑1 system \citep{HuangOrmel2022}. The edge finally settles at the orbit of planet~e. Consequently, the inner planets (b, c1, c2) experience an inward Lindblad torque because they lie between the disk edge and its interior 2:1 commensurability (see Appendix~\ref{sect:LT}). Planet~c2 receives the strongest torque, which breaks its two‑body resonance with e and destabilizes the system (see Fig.~\ref{Fig-SIII_semi}). The following criteria must then be satisfied to match the observed architecture:
\begin{enumerate}
    \item Planets~c1 and~c2 merge into the present‑day planet~c, without any subsequent mergers.
    \item After the merger, the Lindblad torque continues to push planet~c inward. The disk must dissipate when the period ratios are close to the observed values.
\end{enumerate}

We carried out 232 $N$-body simulations, varying the inclination $i$, the mass ratio $m_{\mathrm{c1}}/m_{\mathrm{c}}$, and the nominal Type-I migration timescale $t_{a,\oplus}$ (Table~\ref{Tab2}). Fig.~\ref{Fig-SIII_semi} displays the semi‑major axis and eccentricity evolution for one successful example ($i=45^\circ$, $m_{\mathrm{c1}}/m_{\mathrm{c}}=0.8$). The Lindblad torque drives planet~c2 inward, breaking the c2–e resonance at $\sim$0.75\,Myr. The following instability excites eccentricities and leads to a merger of c1 and c2 at $\sim$1.8\,Myr, forming the new planet~c. Thereafter, the torque continues to push planet~c inward until the period ratios approach the observed configuration at $\sim$2\,Myr, when the disk is assumed to disperse.

Among these 232 simulations, 110 runs ($\sim$47\%) produced a c1--c2 merger, with the resulting four-planet system surviving for at least $10^4$\,yr. After the merger, the planets continue to evolve under the disk torque. {We record the period ratios when $dP$ (Eq.~\ref{eq:SIII-dP}) reaches its minimum, together with the elapsed time from the start of the simulation, $t_{\rm sim}$.}

\begin{figure}
\centering
\includegraphics[width=\textwidth]{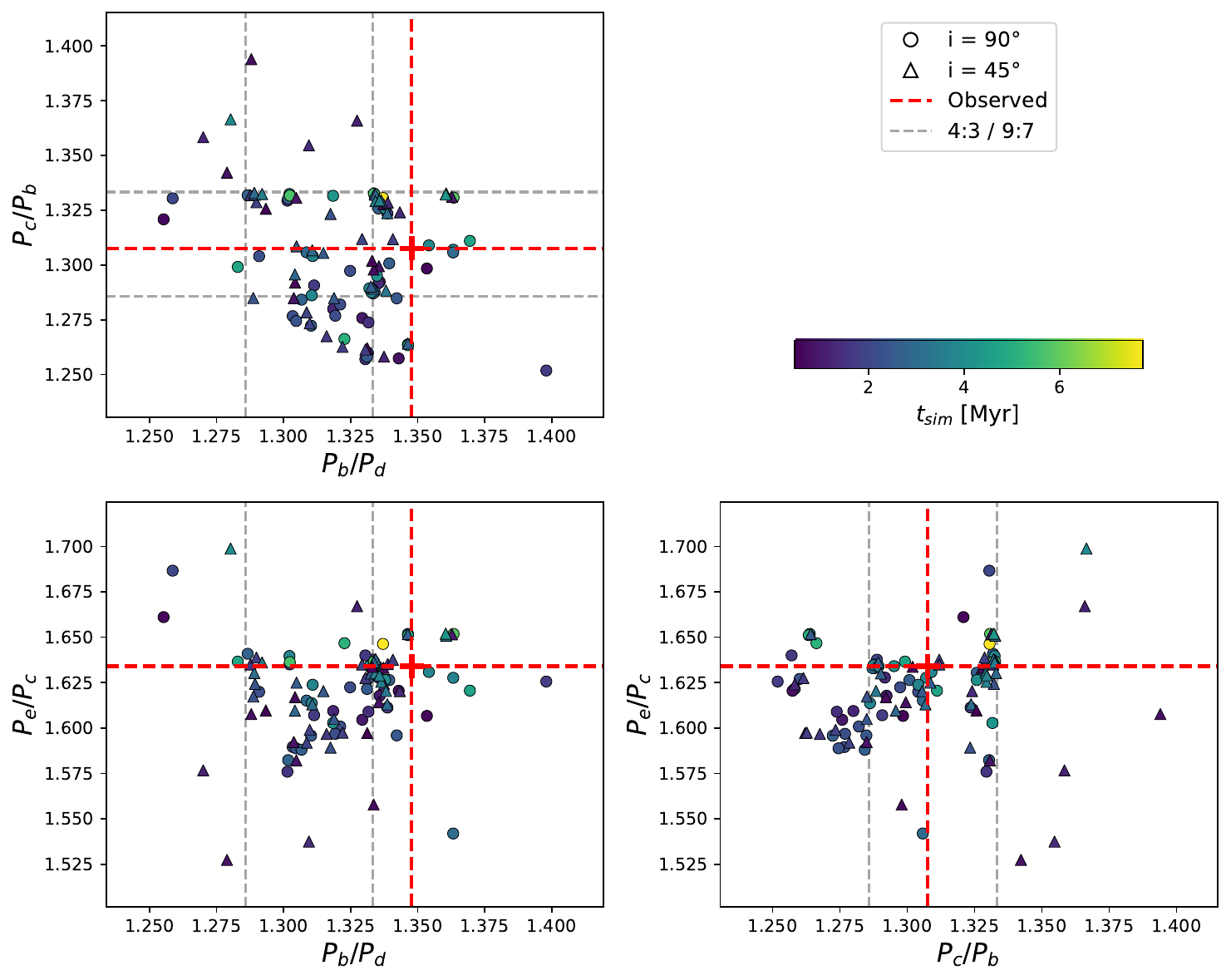}
\caption{Distribution of adjacent planet period ratios for successful merger simulations at minimum $dP$ (Eq.~\ref{eq:SIII-dP}) for Scenario {IIB}. Different colors and symbols show the simulation time $t_{\rm sim}$. The red dashed line indicates the observed period ratio; gray dashed lines mark the 4:3 and 9:7 resonance locations.}
\label{Fig-SIII_dist}
\end{figure}

Fig.~\ref{Fig-SIII_dist} shows the adjacent period ratios at the minimum of $dP$ for all successful mergers. Although the outcome is stochastic and some planet pairs become trapped in two‑body resonances (gray dashed lines), a significant fraction of simulations reach period ratios very close to the observed values, independent of the assumed inclination and other model parameters. Also, the orbital configuration remains stable after disk dispersal because disk damping leaves the planets with small eccentricities. Thus, Scenario~{IIB} can reproduce the orbital architecture of Barnard’s Star provided that the disk disperses at the appropriate epoch.

The color bar in Fig.~\ref{Fig-SIII_dist} indicates simulation time $t_{\rm sim}$, measured from the beginning of the simulation
to the epoch at which the system reaches the configuration closest to the
observed period ratios. This interval includes the Lindblad-torque-driven expansion, the c1--c2 merger, and the subsequent evolution until disk dispersal. The value of $t_{\rm sim}$ is controlled
primarily by the strength of the one-sided Lindblad torque, which scales inversely with the nominal Type-I semi-major axis damping timescale $t_{a,\oplus}$ (see Appendix~\ref{sect:LT}).

Figure~\ref{Fig-SIII_ta} shows this dependence directly. In general, $t_{\rm sim}$ increases with $t_{a,\oplus}$: a longer $t_{\rm sim}$ corresponds to a weaker Lindblad torque and larger $t_{a,\oplus}$. $t_{\rm sim}$ also decreases with planet mass (or smaller inclination $i$), because more massive planets experience stronger Lindblad torque with the same $t_{a,\oplus}$. The relation shows substantial scatter because resonance breaking and planet merging are chaotic. For the successful simulations shown here, $t_{\rm sim}$ spans approximately $0.5$--$8$ Myr. This range is consistent with the characteristic lifetimes of protoplanetary disks \citep{MamajekEric2009,ErcolanoPascucci2017}
(see Appendix~\ref{sect:LT}). The final closeness to the observed configuration, quantified by $dP$, shows no clear dependence on $t_{a,\oplus}$. This suggests that successful simulations may also occur at larger $t_{a,\oplus}$ and correspondingly longer $t_{\rm sim}$.

In summary, Scenario~{IIB} requires the disk inner edge to migrate to the outermost planet or the inner planets to migrate into the disk cavity, and the final period ratios depend sensitively on the disk dissipation time. The merger of c1 and c2 is common ($\sim$47\%), and the mechanism can plausibly produce the observed orbital pattern.

\begin{figure}
\centering
\includegraphics[width=\textwidth]{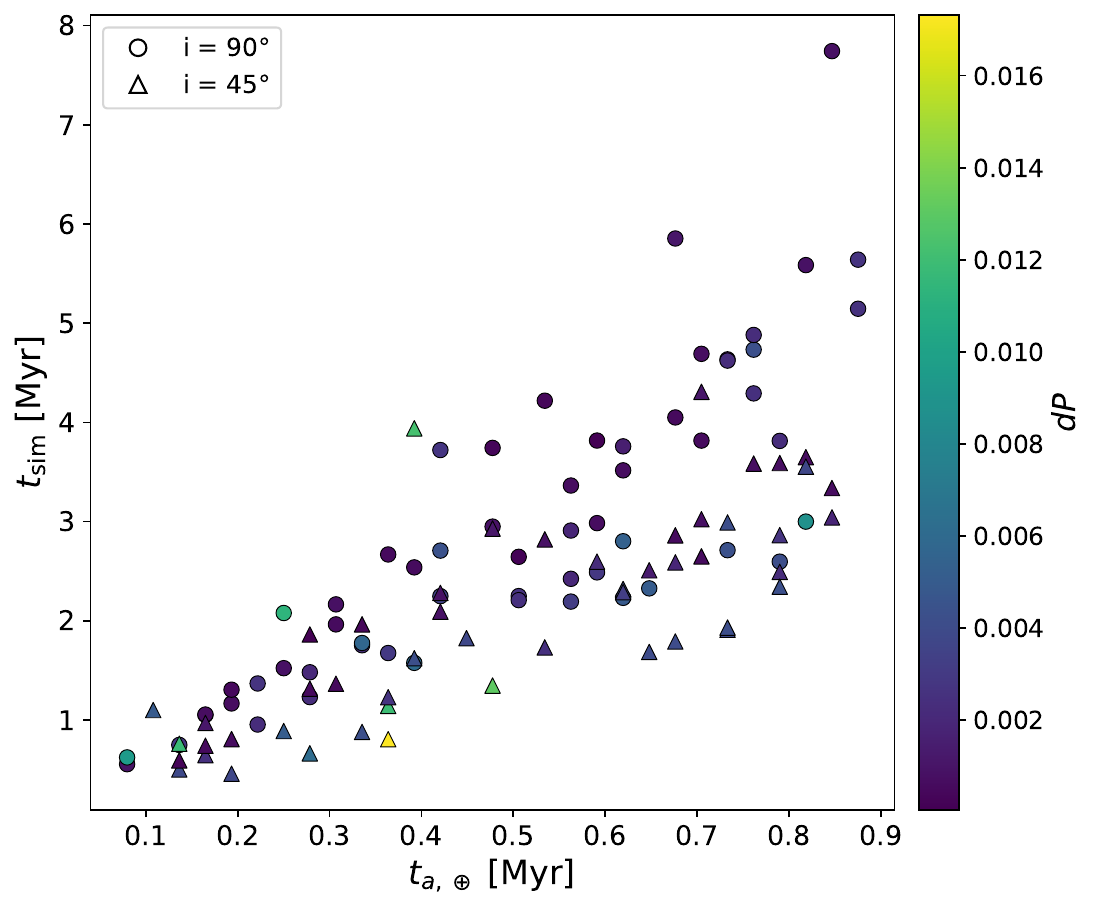}
\caption{Relation between the nominal Type-I semi-major axis damping timescale $t_{a,\oplus}$ adopted in the simulations and the corresponding evolution time $t_{\rm sim}$. Symbol shapes indicate the inclination $i$, corresponding to different planetary masses, and the color bar shows the value of $dP$ in Eq.~\ref{eq:SIII-dP}. Larger $t_{a,\oplus}$ corresponds to a weaker Lindblad torque and generally leads to a longer time for the system to reach the configuration closest to the observed period ratios. }
\label{Fig-SIII_ta}
\end{figure}

\section{Discussion}
\label{sect:discussion}
\subsection{Assessment of the Model}
\label{sect:assessment}

\begin{table*}
\caption{Assessment of the dynamical models for Barnard's Star planetary system.}
\label{tab:model_assessment}
\centering
\small
\setlength{\tabcolsep}{2.5pt}
\begin{tabular}{@{} l p{1.5cm} p{2.5cm} p{4.5cm} p{4.5cm} @{}}
\hline
Scenario & Number of Planets & Resonance-breaking mechanism & Pros & Cons \\
\hline
{IA}   & 4 & None (pure tidal expansion)
    & $\bullet$ Requires neither an additional planet nor a merger event.
    & $\bullet$ The resonance can break before the observed period ratios are reached. \\
    &   & &
    & $\bullet$ The libration amplitude of the $B$-value is too small to encompass the observed offsets (Table~\ref{Tab1}). \\[6pt]
{IB} & \leavevmode{4} & \leavevmode{Magnetospheric rebound}
    & \leavevmode{$\bullet$ Requires neither an additional planet nor a merger event.}
    & \leavevmode{$\bullet$ Planet~e is too low in mass to follow the expanding disk edge efficiently, leaving the outermost period ratio below the observed value in the tested simulations.}

      {$\bullet$ Closer agreement requires additional tuning of the disk dispersal history.} \\[6pt]
\hline
{IIA}  & 5 $\to$ 4 & Tidal expansion after disk dispersal
    & $\bullet$ Tidal expansion naturally moves the resonant chain toward the observed period ratios.

      $\bullet$ Breaking of the 3BR can trigger a c1--c2 merger and produce the present architecture.
    & $\bullet$ A c1--c2 merger occurs in only 12.0\% of the simulations (9/75), and only 6.7\% (5/75) remain stable for the subsequent 10~Myr. \\[6pt]
{IIB} & 5 $\to$ 4 & Lindblad torque during the disk phase
    & $\bullet$ The disk torque efficiently breaks the middle of the chain and produces a c1--c2 merger, with a success rate of 47.4\% (110/232).

      $\bullet$ The preferential disruption of the third planet's resonant links resembles the architecture seen in Barnard's Star, Kepler-221, and many four-planet exoplanetary systems (see Sect.~\ref{sect:application}).
    & $\bullet$ Requires planet~e to end up at the disk inner edge, either because the disk inner edge migrates outward or because the planets are pushed inward.

      $\bullet$ The final period ratios depend sensitively on the epoch of disk dispersal.  \\[6pt]
\hline
\end{tabular}
\end{table*}

We have investigated {four} possible evolutionary histories for Barnard's Star planetary system. {Scenarios~IA and IB begin with four planets in a resonant chain and evolve through tidal expansion and magnetospheric rebound, respectively.} Scenarios~{IIA} and {IIB} instead begin with five planets and a merger between c1 and c2, occurring after or before disk dispersal, respectively.

Table~\ref{tab:model_assessment} summarizes the main advantages and limitations of these models. {Scenarios~IA and IB} start with four planets in a resonance chain. {Neither tidal expansion nor magnetospheric rebound} reproduces the observed configuration within the tested parameter ranges. In comparison, Scenarios~{IIA} and {IIB} both provide viable ``break-the-chain'' pathways. Scenario~{IIA} requires the least tuning during evolution because tidal expansion naturally expands the chain toward the observed configuration. However, only 6.7\% produce a four-planet system that remains stable for 10~Myr. Scenario~{IIB} produces the required merger much more frequently, in 47.4\% of the simulations, but the disk must dissipate near the time when the period ratios are close to the observations.

\subsection{{Distinguishing Scenarios IIA and IIB}}
\label{sect:distinguishing}

{Since both Scenarios~IIA and IIB can reproduce the observed architecture, it is useful to consider additional constraints that might distinguish these histories. Giant impacts can remove a substantial fraction of a planet's atmosphere \citep{KegerreisEtal2020,BierstekerSchlichting2019}. However, both Scenarios~IIA and IIB involve a giant impact. So atmospheric loss alone does not distinguish them. Also, \citet{ByrneEtal2026} find that surviving primordial atmospheres are unlikely on these planets and infer that their cool interiors make outgassing unlikely. Under these conditions, the absence of an atmosphere on planet~c alone cannot distinguish between the two histories. }

{The resonant state of the innermost pair offers another possible way to distinguish the scenarios. In Scenario~IIA, tidal evolution over several Gyr breaks the two-body resonances, while three-body resonances continue the migration of the planets. The inner 4:3 resonant angle circulates in all five post-merger systems that remain stable for 10~Myr, shown by the black circles in Fig.~\ref{Fig-SII_dist}. In Scenario~IIB, the strongest Lindblad torque acts on c2, allowing the inner d--b resonance to survive in some outcomes. Among the 110 simulations shown in Fig.~\ref{Fig-SIII_dist}, 34 (30.9\%) retain libration of the inner 4:3 resonant angle. Therefore, if future radial-velocity measurements can establish libration of the inner resonance, this could favor Scenario~IIB over the stable Scenario~IIA outcomes in our sample.}

\subsection{Application to Other Systems}
\label{sect:application}

Barnard's Star and Kepler-221 \citep{YiEtal2025} share a similar dynamical architecture. In both systems, the proposed initial configuration contains five planets, and the breaking of the resonant chain leads to a merger between the third and fourth planets. This feature motivates a broader search for similar systems exhibiting the ``broken chain'' feature.

We define a ``broken chain'' as an orbital configuration in which one planet is not in resonance with its neighboring planets, while the remaining adjacent or non-adjacent pairs lie close to resonances, with period ratios near integer ratios. For three-planet systems, a configuration in which one pair is resonant while the third planet is not in resonance satisfies the criterion that one planet is ``broken'' from a resonance chain, but it is difficult to distinguish such a configuration from an ordinary resonant pair plus an unrelated planet. Here, we focus on four-planet systems for simplicity; such systems can arise from a merger within an original five-planet resonant chain. Systems with more than four planets satisfying the ``broken chain'' criteria are listed in Appendix~\ref{sect:Broken_mode}.

Fig.~\ref{Fig-broken-c} shows the 14 systems selected by this criterion. Systems containing additional outer planets that are dynamically separated from the compact four-planet chain are also included; these outer companions are omitted from the figure for clarity. Among the 14 systems, the broken planet is the innermost planet in four cases and the third planet in ten cases.

 Ten of the 14 candidate systems exhibit the same location of chain disruption as inferred for Barnard's Star and Kepler-221. Assuming that each of the four planets is equally likely to break from the chain, the probability that the third planet is the broken one in at least 10 of 14 systems is only 0.034\%. This indicates that the concentration at the third planet is unlikely to be a random outcome. This pattern does not prove that mergers similar to these two systems occurred, because other processes may also displace planets from resonance. Nevertheless, it supports the possibility that disruption often occurs near the middle-to-outer part of compact chains. Also, such a feature is consistent with Scenario~{IIB}, in which the Lindblad torque acts most strongly on the planet nearest the disk edge and therefore breaks the resonance between the third and fourth planets.

\begin{figure}
\centering
\includegraphics[width=0.8\textwidth]{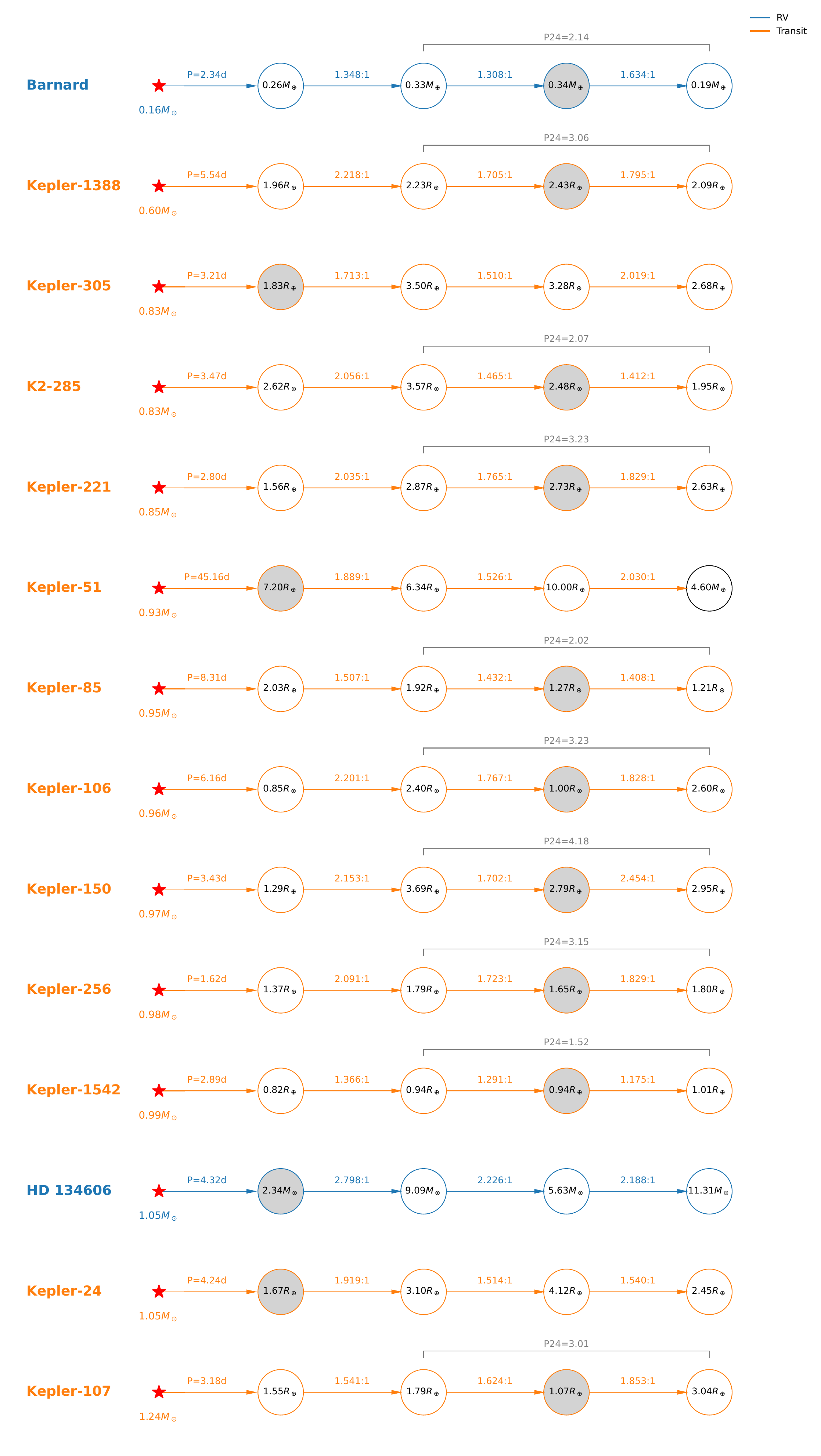}
\caption{Compact planetary systems exhibiting a broken-chain architecture. The inferred broken planet is shown in gray. Systems detected by radial velocity are shown in blue, whereas transit-detected systems are shown in orange. The outermost planet of Kepler-51 was detected only through transit timing variations (TTVs), so its mass, instead of radius, is constrained. For systems with distant outer planets, only the inner four planets are displayed. }
\label{Fig-broken-c}
\end{figure}

\section{Conclusions}
\label{sect:conclusion}

In this work, we investigated {four} possible scenarios for the compact four-planet system around Barnard's Star. Its present architecture resembles a broken resonant chain: the d--b pair lies close to the 4:3 commensurability and the non-adjacent b--e pair lies close to 2:1, whereas planet~c is not in resonance with its adjacent planets. Our simulations disfavor an origin of the observed system in a four-planet resonance chain (Scenarios {IA} and {IB}). On the other hand, the five-planet origin followed by a merger of planets~c1 and~c2 can reproduce the observed configuration. This ``break-the-chain'' event may have occurred either after disk dispersal through long-term tidal expansion (Scenario {IIA}) or during the disk phase under Lindblad torques (Scenario {IIB}).

The main findings of this study are:
\begin{enumerate}
    \item A four-planet resonance chain is unlikely to reproduce the present architecture. In Scenario~{IA}, the four-planet resonant chain broke with tidal expansion before the observed period ratios were reached. Moreover, the libration amplitude of the period ratio remained too small to reach the observed value. {In Scenario~IB, the low mass of the outermost planet~e limits its outward migration through magnetospheric rebound, so reproducing the observed c--e period ratio would require additional tuning.} These results disfavor Scenarios~{IA} and {IB} as the origin of the system.

    \item Planet merging from a five-planet resonance chain could explain the broken chain feature in the system. In Scenario~{IIA}, tidal dissipation expands an initially five-planet resonant chain until the 3BR breaks, triggering the merger of planets~c1 and~c2. In the simulations, 12.0\% produced the desired c1--c2 merger, and 6.7\% remained stable for the subsequent 10~Myr. The main advantage of this scenario is that the tidal evolution both moves the period ratios toward the observed values and naturally breaks the chain, with little tuning required.

    \item The resonant chain may also have broken during the disk phase early on inside the birth disk cavity. In Scenario~{IIB}, planet~e settles at the disk inner edge, and the resulting Lindblad torque acts most strongly on planet~c2. This torque breaks the c2--e resonance and leads to a c1--c2 merger in 110 of the 232 simulations (47.4\%). This pathway is substantially more efficient at producing the required merger than Scenario~{IIA}, but it requires the outermost planet~e to remain at the disk inner edge and the disk to disperse when the outermost period ratio approaches the observed value. {The time required to reach the observed configuration is consistent with the typical few-Myr lifetime of protoplanetary disks (see Appendix~\ref{sect:LT}).}
\end{enumerate}

The dynamical scenarios explored for Barnard's Star can also be applied to other planetary systems with broken resonant chains. We identify 14 compact four-planet systems with broken-chain architectures, in which only one planet is ``broken'' from the resonance chain. In ten systems, the third planet is broken from the chain. The dominance of the third-planet "breaker" is qualitatively consistent with Scenario~{IIB}, in which the Lindblad torque preferentially disrupts resonant links near the middle-to-outer part of the chain.

\begin{acknowledgements}
This work was funded by the National Natural Science Foundation of China (NSFC) under grants No.12473065 and 12233004. The authors gratefully thank Yu Wang, Helong Huang, Zhixuan Li and Xiang Ji for their useful discussions. Software: matplotlib \citep{Hunter2007, CaswellEtal2021}, REBOUND \citep{ReinLiu2012}, REBOUNDx \citep{TamayoEtal2020}. We used ChatGPT (OpenAI) to assist with language editing and the debugging of analysis and plotting scripts. All scientific content,
calculations, and conclusions were independently checked by the authors.
\end{acknowledgements}

\appendix

\section{Lindblad Torque}
\label{sect:LT}

In Scenario~{IIB}, the outermost planet~e is assumed to reside at the inner edge of the gas disk. Near the disk edge, the disk torque becomes one-sided because the gas interior to the truncation radius is depleted. For a low-mass planet at the edge, the positive one-sided corotation torque can balance the negative Lindblad torque and halt inward migration, trapping the planet at the disk inner edge \citep{PaardekooperPapaloizou2009,LiuEtal2017,HuangOrmel2022,RomanovaEtal2026}.

\subsection{Migration Barrier at the Disk Inner Edge}
\label{sect:edge-T}
In this subsection, we investigate the torque on planet e which is at the disk inner edge. We implement the semi-major axis and eccentricity damping of planet~e following the prescription of \citet{HuangOrmel2022}, based on standard disk--planet interaction models \citep{PapaloizouLarwood2000}.

The corresponding damping timescales are \citep{HuangOrmel2022}:
\begin{align}
	t_{a,i} &=
	t_{a,\oplus}
	\left(\frac{q_i}{q_\oplus}\right)^{-1}
	\frac{1}{f_a},
	\label{eq:LT_ta}\\
	t_{e,i} &= C_e h^2 t_{a,i}\frac{f_a}{f_e},
	\label{eq:LT_te}
\end{align}
where $q_i=m_i/m_\star$, $q_\oplus=M_\oplus/m_\star$, $h$ is the disk aspect ratio, $C_e$ is the eccentricity-damping prefactor and $t_{a,\oplus}$ is the nominal Type-I migration timescale (see Eq.~\ref{eq:ta_oplus}). The functions $f_a$ and $f_e$  describe how the damping forces vary near the disk truncation radius $r_c$:
\begin{equation}
f_a =
\begin{cases}
0, & r\leq r_c-\Delta,\\[2pt]
A_a\dfrac{r_c-\Delta-r}{\Delta},
& r_c-\Delta<r<r_c,\\[6pt]
(A_a+1)\dfrac{r-r_c}{\Delta}-A_a,
& r_c\leq r<r_c+\Delta,\\[6pt]
1, & r\geq r_c+\Delta,
\end{cases}
\label{eq:LT_fa}
\end{equation}
and
\begin{equation}
f_e =
\begin{cases}
0, & r<r_c-\Delta,\\[2pt]
\dfrac{r-r_c+\Delta}{\Delta},
& r_c-\Delta<r<r_c,\\[6pt]
1, & r>r_c.
\end{cases}
\label{eq:LT_fe}
\end{equation}
Here, $A_a$ controls the strength of the semi-major axis migration barrier and $\Delta$ sets the width of the transition region. The sign change of $f_a$ across the disk edge represents the transition from inward Type-I migration in the gas disk to the positive one-sided corotation torque at the cavity boundary \citep{LiuEtal2017,HuangOrmel2022}.

In our simulations, we adopt $A_a=50$, $h=0.03$, $C_eh^2=0.01$, and $\Delta=0.06r_c$. The nominal migration timescale (see Sect.~\ref{sect:num}) is varied over $t_{a,\oplus}=8\times10^4$--$9\times10^5\,\mathrm{yr}$, as summarized in Table~\ref{Tab2}. This range is comparable to the $10^5$--$10^6\,\mathrm{yr}$ Type-I migration timescales commonly considered for low-mass planets in disk-migration calculations \citep{OgiharaEtal2015,XuLai2017}. For simplicity, $t_{a,\oplus}$ is held constant within each simulation. In reality, as the disk disperses and the gas surface density decreases, the migration timescale $t_{a,\oplus}$ increases.

\subsection{Lindblad Torque}
\label{sect:app_Lind-T}

While planet~e remains trapped at the disk inner edge, the inner planets located inside the cavity experience a negative Lindblad torque because some of their outer Lindblad resonances still lie within the exterior gas disk. The exact Lindblad resonance condition is \citep{Ward1997}
\begin{equation}
	m\left[\Omega(r)-\Omega_i\right]
	=
	-\kappa(r)\sqrt{1+m^2h^2},
	\label{eq:LT_res_condition}
\end{equation}
where $\Omega_i$ is the planet's orbital frequency, while $\Omega(r)$ and $\kappa(r)$ are the angular and epicyclic frequencies at the resonance location. For a nearly Keplerian disk, $\kappa(r)=\Omega(r)$ and the period corresponding to the $m$-th outer Lindblad resonance satisfies
\begin{equation}
	\frac{P_{L,m}}{P_i}
	\simeq
	1+\frac{\sqrt{1+m^2h^2}}{m}
	\approx \frac{m+1}{m},
	\label{eq:LT_location}
\end{equation}
where $P_{L,m}$ is the period at the resonance location and $P_i$ is the period of the planet. For $h=0.03$, the first few outer Lindblad resonances therefore lie very close to the 2:1, 3:2, 4:3, and 5:4 commensurabilities.

Here, $P_c$ denotes the Keplerian period at the disk inner edge, rather than the orbital period of planet~c. If
\begin{equation}
	\frac{P_{L,m+1}}{P_i}
	<
	\frac{P_c}{P_i}
	<
	\frac{P_{L,m}}{P_i},
	\label{eq:LT_P}
\end{equation}
then the outer Lindblad resonances of orders $1,\ldots,m$ remain in the gas disk, whereas higher-order resonances lie inside the cavity. The planet consequently experiences an inward one-sided Lindblad torque
\begin{equation}
	\Gamma_L = \;\sum_{m\,\in\,\mathrm{disk}} f_m\Gamma_{a,i}
	\label{eq:LT_app}
\end{equation}
where $\Gamma_{a,i}$ is the semi-major axis damping torque for planet $i$ related to $t_{a,\oplus}$.

\begin{figure}
\centering
\includegraphics[width=0.8\textwidth]{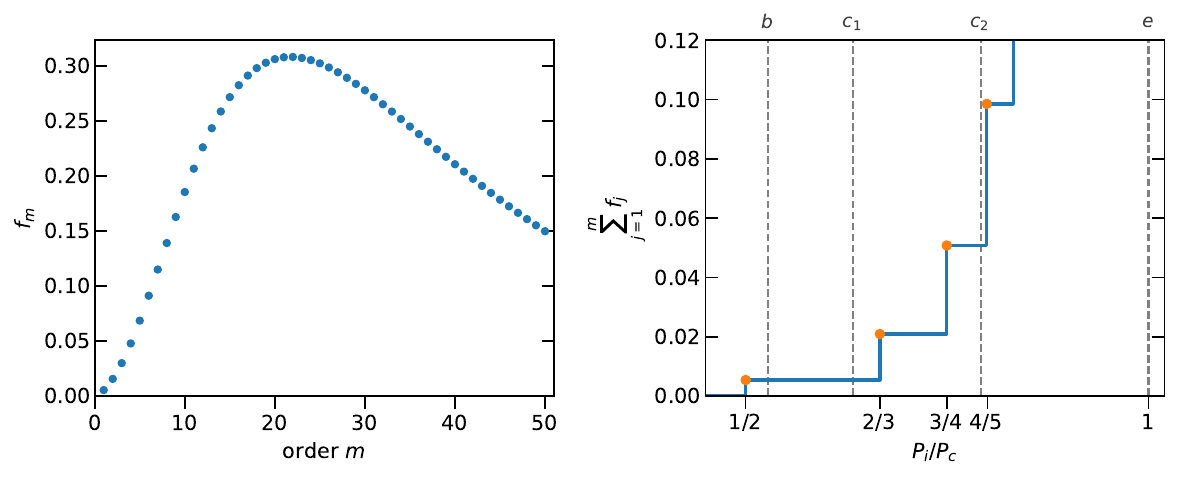}
\caption{Normalized outer Lindblad torque coefficient $f_m$ as a function of resonance order $m$, reproducing Fig. A1 in \citet{HuangOrmel2022} with $h=0.03$, $k=0.5$, and $l=1$. The right panel shows the cumulative torque coefficient as a function of the planet's location relative to the disk inner edge. }
\label{Fig-LT-fm}
\end{figure}

The normalized torque coefficient $f_m$ depends on the disk aspect ratio $h$, the disk surface-density gradient $k$, and the disk temperature gradient $l$ \citep{Ward1997,HuangOrmel2022}. We adopt the same values as \citet{HuangOrmel2022}, $h=0.03$, $k=0.5$, and $l=1$. The resulting coefficients are shown in Fig.~\ref{Fig-LT-fm}.

Because the torque in Eq.~\ref{eq:LT_app} is proportional to the cumulative coefficient $\sum f_m$, planets closer to the disk edge experience a stronger inward torque as additional outer Lindblad resonances remain in the gas. In the initial Scenario~{IIB} configuration (panel~(a) of Fig.~\ref{Fig-SIII}), $P_e/P_{\rm c2}\simeq5/4$, so the first three outer Lindblad resonances can contribute to the torque on planet~c2, giving $\sum_{m=1}^{3}f_m\simeq0.051$ (Fig.~\ref{Fig-LT-fm}). By contrast, planets~b and c1 lie farther inside the cavity and experience weaker Lindblad torques. The enhanced torque on c2 therefore drives more rapid inward migration and results in the breaking of the c2--e resonance and the subsequent instability shown in Fig.~\ref{Fig-SIII_semi}.

Equation~\ref{eq:LT_app} also shows that the one-sided torque scales inversely with $t_{a,\oplus}$. Across the range explored in Table~\ref{Tab2}, increasing $t_{a,\oplus}$ therefore weakens the Lindblad-driven migration and increases the time required for the resonant chain to expand and break.

In Scenario~{IIB}, the disk is assumed to disperse when the post-merger period ratio $P_e/P_c$ approaches the observed value. The resulting evolution typically lasts $\sim0.5$--$8$\,Myr from the onset of Lindblad-driven expansion to the merger and final disk dispersal (see Fig.~\ref{Fig-SIII_dist}). The duration scales approximately with $t_{a,\oplus}$ (see Fig.~\ref{Fig-SIII_ta}), and the range found is compatible with the characteristic lifetime of protoplanetary disks \citep{MamajekEric2009,ErcolanoPascucci2017}.

\section{Switching 3BR Mode}
\label{sect:3BR_mode}

\begin{figure}
\centering
\includegraphics[width=0.8\textwidth]{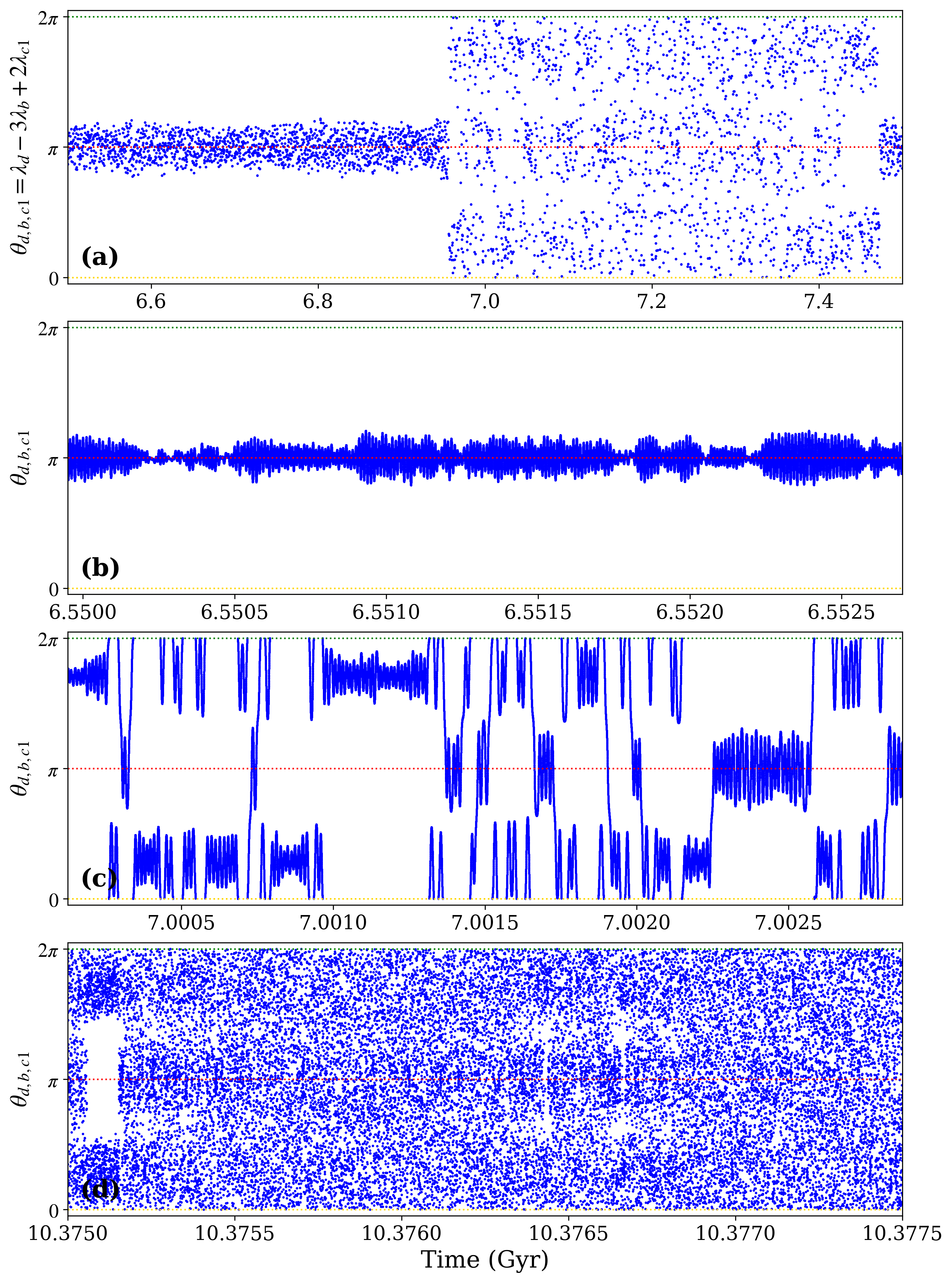}
\caption{Evolution of the three-body resonant angle $\theta_{d,b,c1}=\lambda_d-3\lambda_b+2\lambda_{c1}$ in the Scenario~{IIA} example shown in Fig.~\ref{Fig-SII_R}. Panel~(a) shows the long-term evolution from $6.5$ to $7.5$\,Gyr, during which the angle changes from ordinary libration to an apparent circulating state. Panel~(b) magnifies a short interval during normal libration. Panel~(c) magnifies the apparent circulating phase and reveals rapid switching between different libration centers. Panel~(d) shows a later interval after the 3BR has fully broken around $10$\,Gyr, where the resonant angle fills the entire $0$--$2\pi$ range even on a short timescale.}
\label{Fig-3BR-switch}
\end{figure}

During the tidal expansion in Scenario~{IIA}, we find that three-body resonance can enter a transitional state in which the resonant angle appears to circulate when viewed over long timescales, even though the system has not yet fully left the 3BR. We refer to this behavior as the ``switching 3BR mode''. Fig.~\ref{Fig-3BR-switch} provides a magnified view of the inner 3BR shown in Fig.~\ref{Fig-SII_R}. Panel~(a) shows the evolution of the resonant angle
\begin{equation}
    \theta_{d,b,c1}
    = \lambda_d - 3\lambda_b + 2\lambda_{c1}
\end{equation}
for planets d, b, and c1 between $6.5$ and $7.5$\,Gyr. Before $\sim7$\,Gyr, the angle clearly librates around $\pi$. At later times, the angle appears to circulate around the full interval $0$--$2\pi$. Nevertheless, the period ratios continue to evolve along the 3BR during this phase, as seen in Fig.~\ref{Fig-SII_semi}.

The short-time behavior in panels~(b) and (c) explains this contradiction. Panel~(b) shows a representative interval during the normal libration phase, where $\theta_{d,b,c1}$ remains confined around a single libration center. In panel~(c), the angle is still locally librating, but its libration center repeatedly switches between different values. When sampled over a much longer interval, these rapid switches populate the entire $0$--$2\pi$ range and therefore mimic circulation. Therefore, the planets can remain in 3BR while tidal dissipation continues to expand the chain. Similar changes in three-body resonant libration centers have been reported for the TRAPPIST-1 system \citep{Mah2018}.

This switching state is distinct from a fully broken resonance. Panel~(d) shows the inner 3BR during the later evolution, when $\theta_{d,b,c1}$ continuously fills the full $0$--$2\pi$ interval even on the short timescale shown. At this stage there is no persistent libration center, indicating that the 3BR has broken. This transition results in instability that ultimately leads to the c1--c2 merger in Scenario~{IIA}.

\section{Broken-Chain Systems with More than Four Planets}
\label{sect:Broken_mode}

\begin{figure}
\centering
\includegraphics[width=\textwidth]{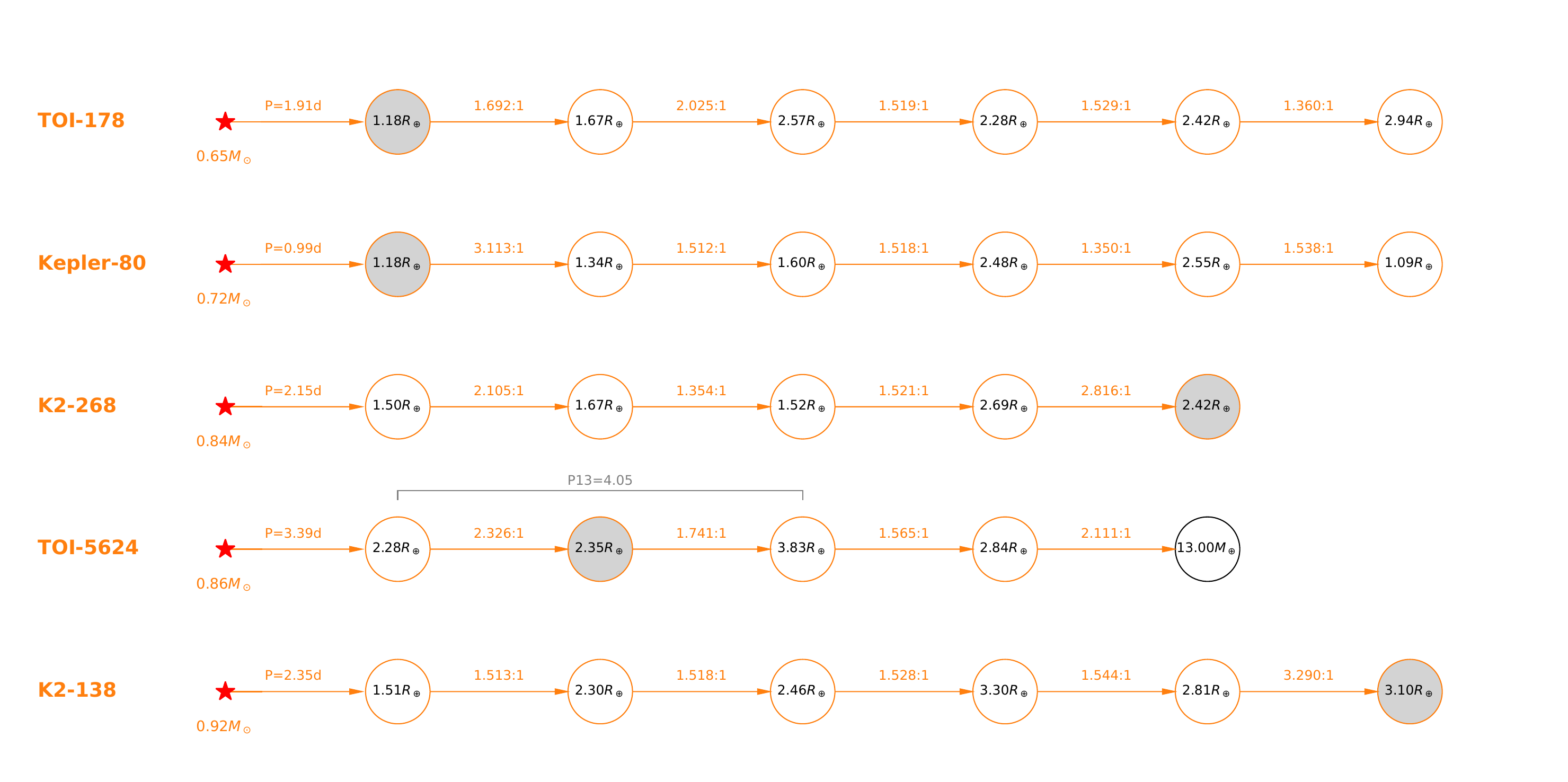}
\caption{Planetary systems with more than four planets satisfying the broken-chain criterion in Sect.~\ref{sect:application}. All systems shown here were discovered through transits. For TOI-5624, the outermost planet was identified only through TTVs and is therefore represented using its mass. }
\label{Fig-more-broken}
\end{figure}

Fig.~\ref{Fig-more-broken} shows systems with more than four planets that satisfy the broken-chain criterion introduced in Sect.~\ref{sect:application}. The orbital architecture of these systems is generally more complex, and satisfying this criterion does not necessarily imply that the system represents a broken resonant chain. For example, in TOI-178 the inner two planets may be trapped in a higher-order resonance while the outer planets form a stable first-order resonant chain. Also, the outer three planets of K2-138 are known to be in a first-order three-body resonance \citep{CerioniBeauge2023}. On the other hand, some higher-multiplicity systems, such as TOI-5624, more closely resemble the broken-chain architecture considered in this work, with one middle planet disrupting a resonant sequence.

\clearpage
\bibliographystyle{raa}
\bibliography{bibtex}

\label{lastpage}

\end{document}